\documentclass[%
reprint,
nofootinbib,
 amsmath,amssymb,
 aps,
]{revtex4-2}

\usepackage{graphicx}
\usepackage{dcolumn}
\usepackage{bm}

\usepackage[T1]{fontenc}

\usepackage{graphicx}
\usepackage{textcomp}
\usepackage{color}
\usepackage{dsfont}
\usepackage{siunitx}
\usepackage{mathtools}
\usepackage{physics}
\usepackage{upgreek}
\usepackage{float}
\usepackage{tikz}
\usepackage{xcolor}
\usepackage{txfonts}
\usepackage[colorlinks=true, linkcolor=black, citecolor=black, urlcolor=blue]{hyperref}

\DeclareSIUnit{\Gauss}{G}
\DeclareSIUnit{\rad}{rad}

\newcommand{\hatnmpk}{|n_{\text{m}, {\km}}|}
\newcommand{\hatnm}{n_{\text{m}, \mathbf{k}}}
\newcommand{\nm}{n_{\text{m}}}
\newcommand{\km}{\mathbf{k}_{-}}
\newcommand{\kplus}{\mathbf{k}_{+}}
\newcommand{\kp}{\mathbf{k}_p}
\newcommand{\kc}{\mathbf{k}_c}
\newcommand{\uc}{u_{\text{c}}}
\newcommand{\up}{u_{\text{p}}}
\newcommand{\R}{\mathbf{r}}

\newcommand{\Erk}{E_{\text{R}}}

\newcommand{\I}{\text{i}}
\newcommand{\E}{\text{e}}
\newcommand{\D}{\text{d}}

\definecolor{brocO_0}{HTML}{9e9e6c}
\definecolor{brocO_1}{HTML}{537099} 

\definecolor{lipari_0}{HTML}{031326} 
\definecolor{lipari_1}{HTML}{bc6461}
\definecolor{lipari_2}{HTML}{e4c1c0} 
\definecolor{lipari_3}{HTML}{d09290}
\definecolor{lipari_4}{HTML}{e7a37a}

\DeclareRobustCommand\sampleline[1]{%
  \tikz\draw[#1, line width=1.8pt] (0,0) (0,\the\dimexpr\fontdimen22\textfont2\relax)
  -- (1.9em,\the\dimexpr\fontdimen22\textfont2\relax);%
}

\begin{document}

\preprint{APS/123-QED}

\title{Microscopy of cavity-induced density-wave ordering in ultracold gases}

\author{Tabea Bühler} \altaffiliation[These authors contributed equally]{ }
\author{Aur\'elien Fabre}\altaffiliation[These authors contributed equally]{ }
\author{Gaia Bolognini}
\author{Zeyang Xue}
\author{Timo Zwettler}
\author{Giulia Del Pace}
\altaffiliation{Current address: Department of Physics and European Laboratory for Nonlinear Spectroscopy (LENS), University of Florence, 50019 Sesto Fiorentino, Italy}
\author{Jean-Philippe Brantut}
\affiliation{Institute of Physics and Center for Quantum Science and Engineering, Ecole Polytechnique F\'ed\'erale de Lausanne, CH-1015 Lausanne, Switzerland}

\date{\today}

\begin{abstract}
We demonstrate high-resolution {\it in-situ} imaging of density-wave ordering induced by cavity-mediated interactions in a unitary Fermi gas. We observe long-range spatial correlations throughout the formation of density waves, both for adiabatic preparation and following a quench, with a pattern controlled by the cavity mode structure. Our single-shot microscopic images together with the real-time readout of the cavity photons provide access to atom-photon correlations. We use this capability to investigate order fluctuations as a function of time following a quench and to directly confirm the correspondence between optical and atomic observables. Our system opens rich perspectives, from local patterning to correlation measurements in long-range interacting quantum gases. 
\end{abstract}

\maketitle

{\it Introduction---}Cavity quantum electrodynamics is one of the most powerful platforms for quantum simulation. Photon exchanges between atoms within the cavity mode produce a tunable long-range interaction which can reach large strength, overcoming other energy scales such as the one given by temperature \cite{Black:2003aa}, kinetic energy \cite{Baumann:2010aa,Klinder:2015aa}, potential energy \cite{ho:2025aa}, atom-atom repulsion \cite{Klinder:2015ab,Landig:2016aa} or Fermi pressure \cite{zhang:2021tr,helson:2023aa} to yield a superradiant phase. Unique to this platform, photons leaking from the cavity provide real-time, weakly destructive information about the dynamics of the ordering process, allowing for detailed investigations of order formation and dynamics \cite{leonard:2017wx,Dogra:2019aa,dreon:2022aa,wu:2023aa,zwettler:2025ab}. Superradiance is accompanied by spatial ordering of the atoms, which can be detected as Bragg peaks using standard time-of-flight images \cite{Baumann:2010aa}. However, both the photons leaking from the cavity and the Bragg peaks represent global observables, incapable of directly revealing order in-situ: the photon signal carries information integrated over the entire mode volume, and time of flight similarly extracts the cloud-averaged phase coherence between neighboring sites \cite{gerbier:2005aa} without direct access to density correlations. Moreover, time-of-flight does not reveal spatial order in incoherent systems such as Mott insulators or thermal gases, and it has strong limitations for Fermi gases close to unitarity due to interactions during the cloud expansion. For these reasons, local imaging and addressing capabilities have been developed for cold atom experiments in optical lattices \cite{Gericke:2008aa,Gemelke:2009aa,Bakr:2009aa,Sherson:2010aa} with transformative impact, from the detection of non-local orders \cite{Endres:2011aa,hilker:2017ab} to the engineering of complex potential landscapes \cite{Preiss:2015ab}. These breakthroughs call for the development of similar techniques for high-resolution observation and addressing at the scale of the self-ordering patterns in cavity-quantum electrodynamics experiments.

\begin{figure}[!ht]
\includegraphics[width=0.48\textwidth]{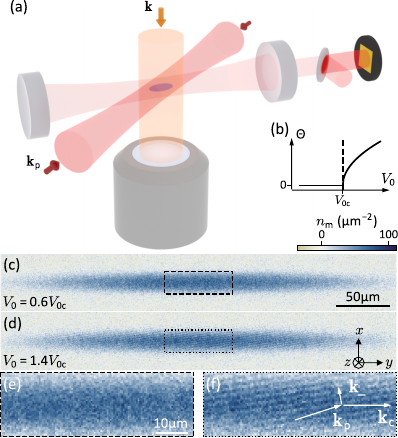}
\caption{Imaging of cavity-induced density waves. (a) Sketch of the experimental setup, with a unitary Fermi gas illuminated by a retroreflected side-pump beam with a wave vector $\textbf{k}_{\text{p}}$ (red). Light scattered by the atoms in the cavity mode (faint red) and leaking through the cavity mirrors is measured using heterodyne detection. A microscope objective oriented along the vertical direction collects absorption images of the atoms, using an imaging beam with wave vector $\textbf{k}$ (light orange). (b) Schematic representation of the phase transition. Above a critical pump strength $V_{0\text{c}}$, the system enters the ordered phase with a finite order parameter $\Theta$. (c)-(f) Single shot absorption images of the atomic cloud for $V_0=0.6V_{0\text{c}}$ (c) and $V_0=1.4V_{0\text{c}}$ (d). (e) and (f) present magnified views on the image centers. In white, the wave vectors of the pump beam ($\textbf{k}_{\text{p}}$), the cavity field ($\textbf{k}_{\text{c}}$) and the resulting difference of the two ($\km$) are illustrated. \label{fig1}}
\end{figure}

The technical overhead of combining high-resolution imaging, quantum degenerate gases and the cavity-QED infrastructure represents a considerable challenge. High-aperture optics and cavity-QED have recently been combined for thermal atoms in optical tweezers applications \cite{dordevic:2021aa,urunuela:2022aa,deist:2022aa,liu:2023aa,zhang:2024aa,hartung:2024aa,hu:2025aa}. Innovative systems combining imaging and cavities along a single optical axis have been demonstrated on thermal gases \cite{orsi:2024ab}. However, none of these setups so far produced an in-situ image of cavity-induced density-wave order. In this paper, we present direct single-shot in-situ absorption imaging of density waves formed in a unitary Fermi gas with strong cavity-mediated interactions. Our system, outlined in Fig.~\ref{fig1}(a), combines the real-time detection of cavity photons used in our previous works \cite{helson:2023aa, zwettler:2025ab, zwettler:2025ac} with a high–numerical aperture ($0.39$) imaging system capable of resolving the density waves on individual in-situ images.

{\it Observation of cavity-induced density-waves---}We investigate cavity-induced self-ordering using our setup featuring a unitary Fermi gas in a high-finesse optical cavity \cite{Roux:2020aa}. We illuminate the atoms with a side-pump laser, which is detuned by $-2\pi \times \SI{47.690}{\GHz}$ from the atomic $D_2$ transition and red-detuned by $\tilde{\Delta}_c \sim \mathrm{MHz}$ from the dispersively shifted cavity resonance. For an applied pump potential $V_0$ exceeding a critical value $V_{0\text{c}}$, a macroscopic density-wave emerges, as illustrated in Fig.~\ref{fig1}(b). Its spatial structure is determined by the interference pattern between the pump and cavity modes, and its amplitude $\Theta$ serves as the order parameter for the phase transition \cite{helson:2023aa}.

Our experiments start with a degenerate two-component unitary Fermi gas comprising typically $4.9(6) \times 10^5$ $^6$Li atoms, at a temperature of $T/T_F \approx 0.16$, with a Fermi energy of $h \times \SI{20(1)}{\kHz}$, in the mode of a high-finesse optical cavity \cite{Buhler:2025}. As it is sketched in Fig.~\ref{fig1}(a), the pump beam is retro-reflected and intersects the cavity at an angle of $\SI{18}{\degree}$ \cite{helson:2023aa,zwettler:2025ab}. $\Theta$ thus comprises two spatial frequency components $\mathbf{k}_{-}$ and $\mathbf{k}_{+}$ with periods of $\SI{2.14}{\micro\meter}$ and $\SI{0.34}{\micro\meter}$, respectively. A commercial aspheric lens is used to image the $(\mathbf{k}_{-},\mathbf{k}_{+})$ plane using absorption imaging \cite{supp}. Critically, the numerical aperture of our lens yields a diffraction limit for coherent imaging of $\lambda_a/\mathrm{NA} = \SI{1.7}{\micro\meter}$, $\lambda_a$ being the wavelength of the imaging light, allowing us to resolve the long-wavelength component of the order parameter at $\mathbf{k}_{-}$. In addition, the cavity field leaking through one of the cavity mirrors is monitored using a heterodyne detection setup as sketched in Fig.~\ref{fig1}(a) \cite{supp}. 

For a given realization of the experiment, we prepare the degenerate atomic gas and then increase adiabatically $V_0$ with a ramp rate of $\dot{V}_0 = 0.05 \times E_\mathrm{R}/\mathrm{ms}$, where $\Erk$ denotes the atomic recoil energy associated to $\textbf{k}_{\text{p}}$, the wave vector of the pump beam, $\Erk = \hbar^2 \textbf{k}_{\text{p}}^2/2m = h \, \times \, \SI{73.67}{\kilo \hertz}$. Upon reaching a target value, we immediately take an absorption image with a measurement duration of \SI{1}{\micro\second}. As it is shown in Fig.~\ref{fig1}, for low pump power (c),(e) the cloud has a Gaussian profile reflecting the underlying trap shape, while above $V_{0\text{c}}$ (d),(f) the atomic density features a modulation at the wave vector $\mathbf{k}_{-}$, a direct manifestation of density-wave order. Since the imaging wavelength is larger than the modulation wavelength associated with $\mathbf{k}_{+}$, the high spatial-frequency component of the density cannot be observed by absorption imaging.

\begin{figure*}[ht]
\centering
\includegraphics[width=1.\textwidth]{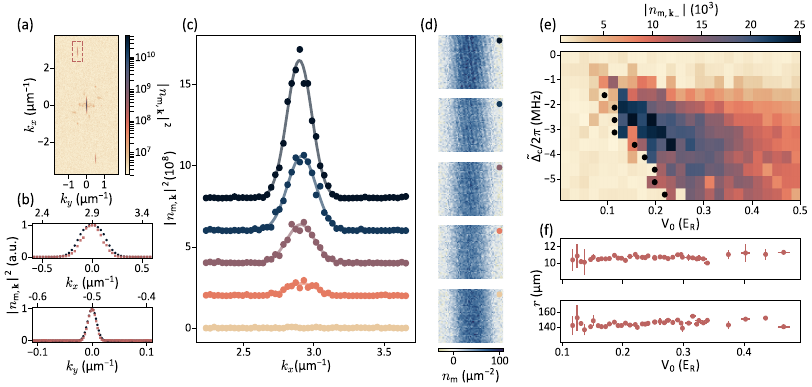}
\caption{Analysis of absorption pictures in Fourier space and imaging across the phase transition. (a) Modulus squared of Fourier transforms of measured atomic densities at an applied pump potential $V_0=1.6(1)V_{0\text{c}}$ (average over 6 repetitions). (b) Line cuts of (a) through the peak at $\mathbf{k}_-$  (\textcolor{lipari_1}{$\blacksquare$}), indicated with a dashed square, along the $x-$direction (top) and the $y-$direction (bottom). In black (\textcolor{lipari_0}{$\medbullet$}) line cuts of (a) at $\mathbf{k}=0$ are shown for comparison. (c) Single-shot line cuts of $|\hatnm|^2$ across the peak at $\mathbf{k}_-$ along $x$ for $V_0=0.6, 1.3, 1.4, 1.5, 1.7V_{0\text{c}}$ (bottom to top, curves vertically offset by 2 $\times$ $10^8$ for clarity). (d) Zoom-in on the cloud centers of the corresponding absorption images in (c). (e) Fourier transform at $\km$ as a function of $V_0$ and $\tilde{\Delta}_c$. The black symbols represent the phase boundary extracted from the onset of the superradiant photon signal. (f) Correlation length of the density wave along the $x$ (top) and $y$-direction (bottom) as a function of $V_0$, above $V_{0\text{c}}$ for $\tilde{\Delta}_{\text{c}}/2\pi =$ $\SI{-2.6(2)}{\MHz}$. The horizontal error bars represent the standard deviation of repeated measurements. The vertical error bars are obtained from a bootstrap resampling analysis.
\label{fig2}
}
\end{figure*}

{\it Image analysis and interpretation---}To analyze the absorption images, we extract the measured column atomic density $\nm$ following the technique of resonant, high-intensity absorption imaging \cite{Reinaudi:2007aa,Horikoshi:2017ab, hueck_calibrating_2017, supp}, and evaluate the two-dimensional Fourier transform of the images $\hatnm$. The modulus squared $|\hatnm|^2$ is presented in Fig.~\ref{fig2}(a), for $V_0 = 1.6(1) V_{0\text{c}}$. Besides the peak around $\mathbf{k}=0$, which reflects the envelope of the atomic cloud, it shows two sharp peaks at the $\pm\km$ points, corresponding to the density modulation visible in real space. Using an independent calibration of the magnification of the imaging system, we extract a period of $\SI{2.14(6)}{\micro\meter}$ of the modulation, which is in agreement with the expected period along $\mathbf{k}_{-}$.
 
Quantitatively interpreting the modulation observed on the images requires understanding of coherent image formation for thick three-dimensional objects. Physically, the density-modulated cloud operates as a thick amplitude diffraction grating, where the first diffraction orders are captured by the imaging system and interfere with the zeroth order on the camera sensor to form the image. Due to the three-dimensional nature of the atomic cloud, the scattering efficiency into the first orders of diffraction is reduced for normal incidence of the imaging beam compared to incidence at the Bragg angle (see \cite{supp} for a quantitative discussion). To realize conditions in which imaging still reveals the density modulation, even for thick atomic clouds, images are taken with the imaging beam deviating from being orthogonal to $\mathbf{k}_{-}$ by an angle of $\alpha \approx \SI{5.2(9)}{\degree}$. In these conditions, we evaluate the suppression factor between the measured Fourier component at $\km$ and the actual Fourier component $n_{\text{m},\km} /n_{\km}$ to be on the order of $0.24(8)$ \cite{supp}. The imaging of the envelope of the atomic cloud remains unaffected.

This effect prevents us from directly interpreting Fig.~\ref{fig2}(a) as the spatial structure factor of the cloud. Nevertheless, in the immediate vicinity of $\mathbf{k}_{-}$, we can interpret $|\hatnm|^2$ in terms of density correlations with a constant suppression factor \cite{supp}. Fig.~\ref{fig2}(b) shows cuts of Fig.~\ref{fig2}(a) along $x$ and $y$ through the peak at $\mathbf{k}_{-}$, indicated with a dashed square, together with the profile around $\mathbf{k} = 0$. Both profiles are normalized to one to cancel the suppression factor, making the shapes directly comparable. The two track each other closely, which is expected if the envelope of the density modulation follows the slowly varying density profile of the cloud. Our observations support the expectation that the density modulation has a single spatial frequency and is coherent over the entire atomic cloud. This is further confirmed by a local analysis of both Fourier amplitude and phase at $\mathbf{k}_{-}$, showing a uniform phase over the full extent of the atomic cloud \cite{supp}.

{\it Imaging across the transition---}The Fourier spectrum around $\mathbf{k}_{-}$ allows us to track the density-wave order as it emerges above $V_{0\text{c}}$. Fig.~\ref{fig2}(c) presents cuts along the $x-$direction of $|\hatnm|^2$ through the peak at $\mathbf{k}_{-}$ as $V_0$ is increased across $V_{0\text{c}}$, showing the onset and growth of density-wave order. Zooms on the center of the corresponding real-space images are shown in Fig.~\ref{fig2}(d). Extracting the amplitude of the Fourier peak at $\km$ $\hatnmpk$ via a Gaussian fit as a function of different system parameters allows to reconstruct the phase diagram in the $\tilde{\Delta}_c$-$V_0$ plane, as it is shown in  Fig.~\ref{fig2}(e), where $V_0$ is given in units of $\Erk$. The phase boundary is faithfully reconstructed. Independently, we collect photons emerging from the cavity as $V_0$ is ramped up, allowing us to track the emergence of density-wave through its superradiant character, as it was done in all previous cavity-induced ordering experiments including our previous work \cite{Baumann:2010aa,zhang:2021tr,helson:2023aa}. The phase boundary extracted from the photon signal is indicated in Fig.~\ref{fig2}(e) as black symbols (see \cite{supp} for more details) and is in good agreement with the one extracted from the absorption images.  

The width of the peak at $\mathbf{k}_{-}$ in $|\hatnm|^2$ can be interpreted as the inverse of the correlation length $r$ of the density wave. We use a Gaussian fit to track $r$ as $V_0$ is increased. The results are shown in Fig.~\ref{fig2}(f) for extraction along the $x$ and $y$ direction, for $\tilde{\Delta}_{\text{c}}/2\pi =$ $\SI{-2.6(2)}{\MHz}$, showing almost no dependence on $V_0$. The extracted correlation lengths in the two directions differ by an order of magnitude, which reflects the aspect ratio of the atomic cloud. Closest to the critical point, error bars are dominated by shot-to-shot fluctuations of the exact threshold. However, for every data point showing a finite density-wave order the width of the Fourier peak is consistent with the width at higher $V_0$ within error bars. This demonstrates that the buildup of order occurs uniformly across the cloud, which is the expected behavior of ordering driven by infinite-range interactions \cite{marijanovic:2024aa,zwettler:2025ab} and is in agreement with the single-mode character of the cavity. It contrasts with order growing from local interactions, where a fluctuating regime with limited spatial correlations is expected to exist in the vicinity of the transition \cite{sethna2021statistical}. 

\begin{figure}[t]
\includegraphics[width=0.48\textwidth]{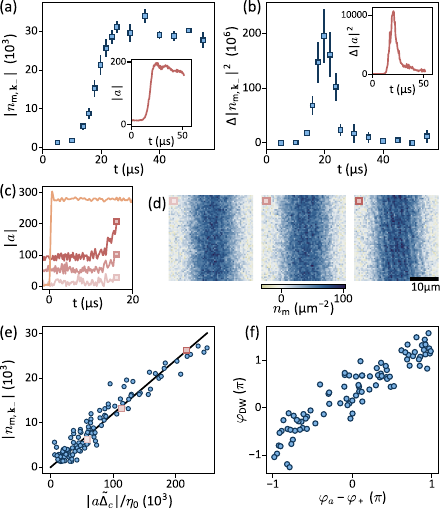} 
\caption{Density-wave ordering after a quench. (a) Mean density modulation amplitude as a function of time, following a quench to $V_0 = 14.4(4)V_{0\text{c}}$. Inset: mean cavity field amplitude for the same realizations. The vertical error bars represent the standard deviation of repeated measurements. (b) Variance of the density modulation amplitude. Inset: variance of the cavity field amplitudes. The vertical error bars represent the statistical uncertainty due to finite sample size. (c) Cavity field amplitude for three experimental realizations at time $t=\SI{16}{\micro\second}$, (\textcolor{lipari_2}{\sampleline{}}, \textcolor{lipari_3}{\sampleline{}}, \textcolor{lipari_1}{\sampleline{}}) (offset for clarity). The extracted field endpoints are shown as a square symbols. The pump field is represented by the orange line (\textcolor{lipari_4}{\sampleline{}}). (d) Zoom-in on absorption images recorded at the end of the three experimental realizations shown in (c). (e) Correlation between density-wave modulation and cavity field amplitude, each data point representing an individual experimental realization. The solid line represents a linear fit to the data. The realizations from panels (c)-(d) are marked as square symbols. (f) Correlation between the phase of the density-wave and the phase of the cavity field referenced to the phase of the pump field.
\label{fig3}}
\end{figure}

{\it Quench dynamics and correlations---}In-situ imaging provides access to density-density correlations in real space. The settings of cavity QED also provide real-time, weakly-destructive information in the form of photons leaking continuously through the cavity mirrors. 
We now use this feature to investigate heterogeneous correlations between atomic and photonic observables, which underpin all the effective models of cavity-induced interactions \cite{mivehvar:2021aa}. For this, we perform an instantaneous quench into the ordered phase \cite{wu:2023aa,zwettler:2025ab} and follow the build-up of order over an ensemble of realizations. 

First, we observe $\hatnmpk$ grow exponentially as the density wave instability builds up, and then saturate. As shown in previous works \cite{wu:2023aa,zwettler:2025ab}, the growth rate can span several orders of magnitude, controlled at a given detuning by the ratio $V_0/V_{0\text{c}}$. The ensemble average of $\hatnmpk$ is shown as a function of time in Fig.~\ref{fig3}(a) for $V_0 = 14.4(4)V_{0\text{c}}$. Its variance $\Delta \hatnmpk ^2$ increases and peaks as the system approaches the steady state, as shown in Fig.~\ref{fig3}(b). Examples of images taken after \SI{16}{\micro\second} illustrate this variability (see Fig.~\ref{fig3}(d)). At longer time, order settles to a finite value and the variance is strongly reduced as the system reaches a steady state. 

We leverage the intrinsic variability of $\hatnmpk$ in the quench experiment to analyze correlations with the cavity photons. On the same ensemble of realizations, we evaluate the average of the cavity field amplitude $|a|$ and its variance $\Delta |a|^2$ as a function of time, as presented in the insets of Fig.~\ref{fig3}(a) and (b). Similar to $\Delta \hatnmpk ^2$, the field variance peaks as the system approaches the steady state and then settles for longer times. Three examples of detected cavity field amplitudes are shown in Fig.~\ref{fig3}(c) for the same experimental realizations as in Fig.~\ref{fig3}(d). The extracted field endpoints are marked as square symbols and the orange line shows the applied $V_0$. We now evaluate density-field correlations over the full set of measurements. Fig.~\ref{fig3}(e) presents the correlation between $\hatnmpk$ and the field amplitude $|a|$ from single realizations, in the early time following the quench. The field amplitude is normalized with the parameter $\eta_0 \propto \sqrt{V_0}$ \cite{supp}. The linear correlation is very strong down to the noise floor. A linear fit is shown as a black line. 

The fluctuations observed in both $\hatnmpk$ and $|a|$ are much larger than the underlying noise in the control parameters. They are expected from the initial quantum or thermal seed from which the density-wave builds \cite{marijanovic:2024aa}, or from dissipation leading to self-organized states with an incoherent mixture of cavity amplitudes \cite{halati:2020aa}. Strong atom-field correlations are expected from the theory of cavity-mediated ordering: for large ${\tilde{\Delta}_c}$, the atomic and photonic degrees of freedom are adiabatically following each other, leading to the operator relation $\hat{a} =\frac{\eta_0}{\tilde{\Delta}_c} \hat{\Theta}$. This relation underlies the description of the system in terms of long-range interacting atoms and has, to our knowledge, never been tested experimentally. To highlight the general character of these correlations, we further confirmed them for quantum gases prepared away from unitarity \cite{supp}. Our observations of large and correlated fluctuations rule out a mean-field description of the atom-light system as a product state $\hat{\rho} = \hat{\rho}_\text{at}\otimes \ket{\alpha(t)}\bra{\alpha(t)}$ between an atomic state $\hat{\rho}_\text{at}$ and a coherent state of the cavity field. Indeed, this would imply that the photon number follows Poisson statistics, i.e. that the variance and mean amplitudes would be proportional, in obvious contradiction with our observations. Instead, our data can be described as an incoherent mixture of coherent states, each of them corresponding to a particular atomic configuration, as conjectured in \cite{halati:2020aa,halati:2020ab}. 

In addition to the cavity field amplitude we measure the phase of the cavity field with respect to the pump beam $\varphi_{\text{a}}-\varphi_{\text{+}}$ \cite{supp}. Its counterpart on the images is the phase of the density modulation referenced to the position of the camera $\varphi_{\text{DW}}$, which can be extracted locally and over individual realizations. This information goes beyond the density correlations measured by the structure factor and can only be accessed by direct in-situ observation. We evaluate phase correlations between the atomic density wave and the cavity field. The results are presented in Fig.~\ref{fig3}(f). Due to the dependence of $\varphi_{\text{DW}}$ on the phase of the pump field (see \cite{supp} for details), its measured values explore the full $2\pi$-range. The correlations between $\varphi_{\text{DW}}$ and $\varphi_{\text{a}}-\varphi_{\text{+}}$, however, are again very strong, demonstrating the stability of the microscope setup throughout several experimental repetitions.

\begin{figure}[t]
\includegraphics[width=0.5\textwidth]{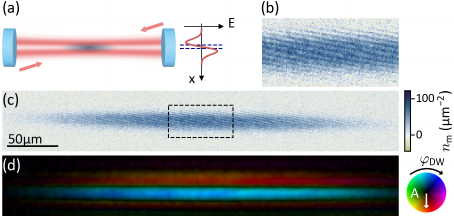}
\caption{Density-wave ordering at the first transverse cavity mode TEM$_{01}$. (a) Sketch of the experiment geometry with the transverse profile of the electric field amplitude of the TEM$_{01}$ mode (red curve) and the full width at half maximum of the atomic density profile (dashed lines). (b) Zoom-in at the center and (c) wide field image of the atomic density for density-wave ordering close to the first transverse mode of the cavity. (d) Local map of the amplitude (brightness) and phase (color map) of the density wave pattern across the cloud, averaged over 100 repetitions.
\label{fig4}}
\end{figure}

{\it Density-wave ordering at higher-order modes---}Our observation of a uniform phase of the density wave in the entire cloud reflects the phase of the underlying TEM$_{00}$ cavity mode. If higher-order modes are involved in the ordering process, the density wave pattern within the atomic cloud can exhibit phase variations that cannot be retrieved by analyzing the overall phase of the emitted light. To demonstrate the ability to image inhomogeneous phase patterns, we reproduced the experiments with the pump laser frequency tuned to the close vicinity of the TEM$_{01}$ cavity mode, located \SI{2.634(1)}{\giga\hertz} away from the TEM$_{00}$ mode. By monitoring the total number of photons leaking from the cavity, we observe the superradiant transition indicating density-wave ordering in this configuration. Exploiting the astigmatism of the cavity, which spectrally separates the two near-degenerate TEM$_{01}$ and TEM$_{10}$ modes \cite{supp}, we target specifically the mode oriented orthogonal to the microscope axis and take images in the ordered phase. A sketch of the experiment is shown in Fig.~\ref{fig4}(a), where the cavity field follows the shape of the TEM$_{01}$ mode, whose electric field along the $x-$ direction is indicated in red. It has a nodal line as well as a $\pi$ phase jump at the cavity axis. An example of an absorption image in this configuration is shown in Fig.~\ref{fig4}(c), with a zoom-in on the image center in Fig.~\ref{fig4}(b). While the envelope of the density distribution is not affected by the change of the cavity mode, the density wave pattern exhibits a phase distortion at the center of the cloud, the location of the optical axis of the cavity. We analyze the recorded absorption images by computing $n_{\text{m}, \km}$ with a spatial resolution of $\SI{3.4}{\micro\meter}$, allowing us to estimate both the amplitude and the phase of the density modulation locally. The results of the extracted amplitude and phase, averaged over $100$ experimental repetitions, are shown in Fig.~\ref{fig4}(d). A $\pi$ phase shift between the two sides of the cloud separated by a nodal line in the amplitude is clearly visible, matching the expected profile dictated by the TEM$_{01}$ cavity mode.

{\it Conclusion and outlook---}We have demonstrated the direct observation of density-wave ordering in an ultracold Fermi gas using resonant absorption imaging, which enables the analysis of density correlations throughout the ordering process, as well as during ordering at higher-order cavity modes. The combination of single-shot imaging with single-trajectory readout of the cavity photons allowed us to reconstruct atom-field correlations, testing for the first time the foundations of cavity-induced interactions. 

The ability to directly image the density-wave order opens a broad range of possibilities. First, it could be directly generalized to phase contrast imaging capable of observing magnetization patterns \cite{Partridge:2006fk,Shin:2006aa}, that could emerge in the case of spin-imbalanced Fermi gases. In addition, imaging combined with radio-frequency spectroscopy \cite{Shin:2007aa} can be used to explore the fate of pairing in the presence of long-range interactions. This would be crucial when operating the experiment close to a photo-association line, where the ordered phase acquires a pair-density wave character \cite{zwettler:2025ac}. 
The microscope objective could also be used to project spatially engineered pump beams translating into complex, on-demand interaction patterns. Together with time-dependent manipulations, this can lead to effective finite-range interactions and new phases of matter \cite{bonifacio:2024aa,baumgartner:2024aa}.

{\it Acknowledgements---}We thank Kyuhwan Lee for careful reading of the manuscript and for discussions, Catalin-Mihai Halati for discussions, and Anastasiia Barysheva and Benedikt Röcken for their contributions to the characterization and mounting design of the imaging system. We acknowledge funding from the Swiss State Secretariat for Education, Research and Innovation (Grants No. MB22.00063 and UeM019-5.1) and the Swiss National Science Foundation (Grant No. 200020E\_217124). A.F. acknowledges funding from the EPFL Center for Quantum Science and Engineering.

{\it Data availability---}The data that support the findings of this article are openly available \cite{data}.

\bibliography{paper_bibliography}

\begin{thebibliography}{55}%
\makeatletter
\providecommand \@ifxundefined [1]{%
 \@ifx{#1\undefined}
}%
\providecommand \@ifnum [1]{%
 \ifnum #1\expandafter \@firstoftwo
 \else \expandafter \@secondoftwo
 \fi
}%
\providecommand \@ifx [1]{%
 \ifx #1\expandafter \@firstoftwo
 \else \expandafter \@secondoftwo
 \fi
}%
\providecommand \natexlab [1]{#1}%
\providecommand \enquote  [1]{``#1''}%
\providecommand \bibnamefont  [1]{#1}%
\providecommand \bibfnamefont [1]{#1}%
\providecommand \citenamefont [1]{#1}%
\providecommand \href@noop [0]{\@secondoftwo}%
\providecommand \href [0]{\begingroup \@sanitize@url \@href}%
\providecommand \@href[1]{\@@startlink{#1}\@@href}%
\providecommand \@@href[1]{\endgroup#1\@@endlink}%
\providecommand \@sanitize@url [0]{\catcode `\\12\catcode `\$12\catcode `\&12\catcode `\#12\catcode `\^12\catcode `\_12\catcode `\%12\relax}%
\providecommand \@@startlink[1]{}%
\providecommand \@@endlink[0]{}%
\providecommand \url  [0]{\begingroup\@sanitize@url \@url }%
\providecommand \@url [1]{\endgroup\@href {#1}{\urlprefix }}%
\providecommand \urlprefix  [0]{URL }%
\providecommand \Eprint [0]{\href }%
\providecommand \doibase [0]{https://doi.org/}%
\providecommand \selectlanguage [0]{\@gobble}%
\providecommand \bibinfo  [0]{\@secondoftwo}%
\providecommand \bibfield  [0]{\@secondoftwo}%
\providecommand \translation [1]{[#1]}%
\providecommand \BibitemOpen [0]{}%
\providecommand \bibitemStop [0]{}%
\providecommand \bibitemNoStop [0]{.\EOS\space}%
\providecommand \EOS [0]{\spacefactor3000\relax}%
\providecommand \BibitemShut  [1]{\csname bibitem#1\endcsname}%
\let\auto@bib@innerbib\@empty
\bibitem [{\citenamefont {Black}\ \emph {et~al.}(2003)\citenamefont {Black}, \citenamefont {Chan},\ and\ \citenamefont {Vuleti\ifmmode~\acute{c}\else \'{c}\fi{}}}]{Black:2003aa}%
  \BibitemOpen
  \bibfield  {author} {\bibinfo {author} {\bibfnamefont {A.~T.}\ \bibnamefont {Black}}, \bibinfo {author} {\bibfnamefont {H.~W.}\ \bibnamefont {Chan}},\ and\ \bibinfo {author} {\bibfnamefont {V.}~\bibnamefont {Vuleti\ifmmode~\acute{c}\else \'{c}\fi{}}},\ }\bibfield  {title} {\bibinfo {title} {Observation of collective friction forces due to spatial self-organization of atoms: From rayleigh to bragg scattering},\ }\href {https://doi.org/10.1103/PhysRevLett.91.203001} {\bibfield  {journal} {\bibinfo  {journal} {Phys. Rev. Lett.}\ }\textbf {\bibinfo {volume} {91}},\ \bibinfo {pages} {203001} (\bibinfo {year} {2003})}\BibitemShut {NoStop}%
\bibitem [{\citenamefont {{Baumann}}\ \emph {et~al.}(2010)\citenamefont {{Baumann}}, \citenamefont {{Guerlin}}, \citenamefont {{Brennecke}},\ and\ \citenamefont {{Esslinger}}}]{Baumann:2010aa}%
  \BibitemOpen
  \bibfield  {author} {\bibinfo {author} {\bibfnamefont {K.}~\bibnamefont {{Baumann}}}, \bibinfo {author} {\bibfnamefont {C.}~\bibnamefont {{Guerlin}}}, \bibinfo {author} {\bibfnamefont {F.}~\bibnamefont {{Brennecke}}},\ and\ \bibinfo {author} {\bibfnamefont {T.}~\bibnamefont {{Esslinger}}},\ }\bibfield  {title} {\bibinfo {title} {{Dicke quantum phase transition with a superfluid gas in an optical cavity}},\ }\href {https://doi.org/https://doi.org/10.48550/arXiv.0912.3261} {\bibfield  {journal} {\bibinfo  {journal} {Nature}\ }\textbf {\bibinfo {volume} {464}},\ \bibinfo {pages} {1301} (\bibinfo {year} {2010})}\BibitemShut {NoStop}%
\bibitem [{\citenamefont {Klinder}\ \emph {et~al.}(2015{\natexlab{a}})\citenamefont {Klinder}, \citenamefont {Ke{\ss}ler}, \citenamefont {Wolke}, \citenamefont {Mathey},\ and\ \citenamefont {Hemmerich}}]{Klinder:2015aa}%
  \BibitemOpen
  \bibfield  {author} {\bibinfo {author} {\bibfnamefont {J.}~\bibnamefont {Klinder}}, \bibinfo {author} {\bibfnamefont {H.}~\bibnamefont {Ke{\ss}ler}}, \bibinfo {author} {\bibfnamefont {M.}~\bibnamefont {Wolke}}, \bibinfo {author} {\bibfnamefont {L.}~\bibnamefont {Mathey}},\ and\ \bibinfo {author} {\bibfnamefont {A.}~\bibnamefont {Hemmerich}},\ }\bibfield  {title} {\bibinfo {title} {Dynamical phase transition in the open dicke model},\ }\href {https://doi.org/https://doi.org/10.48550/arXiv.1409.1945} {\bibfield  {journal} {\bibinfo  {journal} {Proceedings of the National Academy of Sciences}\ }\textbf {\bibinfo {volume} {112}},\ \bibinfo {pages} {3290} (\bibinfo {year} {2015}{\natexlab{a}})}\BibitemShut {NoStop}%
\bibitem [{\citenamefont {Ho}\ \emph {et~al.}(2025)\citenamefont {Ho}, \citenamefont {Lu}, \citenamefont {Xiang}, \citenamefont {Rusconi}, \citenamefont {Masson}, \citenamefont {Asenjo-Garcia}, \citenamefont {Yan},\ and\ \citenamefont {Stamper-Kurn}}]{ho:2025aa}%
  \BibitemOpen
  \bibfield  {author} {\bibinfo {author} {\bibfnamefont {J.}~\bibnamefont {Ho}}, \bibinfo {author} {\bibfnamefont {Y.-H.}\ \bibnamefont {Lu}}, \bibinfo {author} {\bibfnamefont {T.}~\bibnamefont {Xiang}}, \bibinfo {author} {\bibfnamefont {C.~C.}\ \bibnamefont {Rusconi}}, \bibinfo {author} {\bibfnamefont {S.~J.}\ \bibnamefont {Masson}}, \bibinfo {author} {\bibfnamefont {A.}~\bibnamefont {Asenjo-Garcia}}, \bibinfo {author} {\bibfnamefont {Z.}~\bibnamefont {Yan}},\ and\ \bibinfo {author} {\bibfnamefont {D.~M.}\ \bibnamefont {Stamper-Kurn}},\ }\bibfield  {title} {\bibinfo {title} {Optomechanical self-organization in a mesoscopic atom array},\ }\href {https://doi.org/https://doi.org/10.48550/arXiv.2410.12754} {\bibfield  {journal} {\bibinfo  {journal} {Nature Physics}\ }\textbf {\bibinfo {volume} {21}},\ \bibinfo {pages} {1071} (\bibinfo {year} {2025})}\BibitemShut {NoStop}%
\bibitem [{\citenamefont {Klinder}\ \emph {et~al.}(2015{\natexlab{b}})\citenamefont {Klinder}, \citenamefont {Ke\ss{}ler}, \citenamefont {Bakhtiari}, \citenamefont {Thorwart},\ and\ \citenamefont {Hemmerich}}]{Klinder:2015ab}%
  \BibitemOpen
  \bibfield  {author} {\bibinfo {author} {\bibfnamefont {J.}~\bibnamefont {Klinder}}, \bibinfo {author} {\bibfnamefont {H.}~\bibnamefont {Ke\ss{}ler}}, \bibinfo {author} {\bibfnamefont {M.~R.}\ \bibnamefont {Bakhtiari}}, \bibinfo {author} {\bibfnamefont {M.}~\bibnamefont {Thorwart}},\ and\ \bibinfo {author} {\bibfnamefont {A.}~\bibnamefont {Hemmerich}},\ }\bibfield  {title} {\bibinfo {title} {Observation of a superradiant mott insulator in the dicke-hubbard model},\ }\href {https://doi.org/https://doi.org/10.1103/PhysRevLett.115.230403} {\bibfield  {journal} {\bibinfo  {journal} {Phys. Rev. Lett.}\ }\textbf {\bibinfo {volume} {115}},\ \bibinfo {pages} {230403} (\bibinfo {year} {2015}{\natexlab{b}})}\BibitemShut {NoStop}%
\bibitem [{\citenamefont {Landig}\ \emph {et~al.}(2016)\citenamefont {Landig}, \citenamefont {Hruby}, \citenamefont {Dogra}, \citenamefont {Landini}, \citenamefont {Mottl}, \citenamefont {Donner},\ and\ \citenamefont {Esslinger}}]{Landig:2016aa}%
  \BibitemOpen
  \bibfield  {author} {\bibinfo {author} {\bibfnamefont {R.}~\bibnamefont {Landig}}, \bibinfo {author} {\bibfnamefont {L.}~\bibnamefont {Hruby}}, \bibinfo {author} {\bibfnamefont {N.}~\bibnamefont {Dogra}}, \bibinfo {author} {\bibfnamefont {M.}~\bibnamefont {Landini}}, \bibinfo {author} {\bibfnamefont {R.}~\bibnamefont {Mottl}}, \bibinfo {author} {\bibfnamefont {T.}~\bibnamefont {Donner}},\ and\ \bibinfo {author} {\bibfnamefont {T.}~\bibnamefont {Esslinger}},\ }\bibfield  {title} {\bibinfo {title} {Quantum phases from competing short- and long-range interactions in an optical lattice},\ }\href {https://doi.org/https://doi.org/10.1038/nature17409} {\bibfield  {journal} {\bibinfo  {journal} {Nature}\ }\textbf {\bibinfo {volume} {532}},\ \bibinfo {pages} {476} (\bibinfo {year} {2016})}\BibitemShut {NoStop}%
\bibitem [{\citenamefont {Zhang}\ \emph {et~al.}(2021)\citenamefont {Zhang}, \citenamefont {Chen}, \citenamefont {Wu}, \citenamefont {Wang}, \citenamefont {Fan}, \citenamefont {Deng},\ and\ \citenamefont {Wu}}]{zhang:2021tr}%
  \BibitemOpen
  \bibfield  {author} {\bibinfo {author} {\bibfnamefont {X.}~\bibnamefont {Zhang}}, \bibinfo {author} {\bibfnamefont {Y.}~\bibnamefont {Chen}}, \bibinfo {author} {\bibfnamefont {Z.}~\bibnamefont {Wu}}, \bibinfo {author} {\bibfnamefont {J.}~\bibnamefont {Wang}}, \bibinfo {author} {\bibfnamefont {J.}~\bibnamefont {Fan}}, \bibinfo {author} {\bibfnamefont {S.}~\bibnamefont {Deng}},\ and\ \bibinfo {author} {\bibfnamefont {H.}~\bibnamefont {Wu}},\ }\bibfield  {title} {\bibinfo {title} {Observation of a superradiant quantum phase transition in an intracavity degenerate fermi gas},\ }\href {https://doi.org/10.1126/science.abd4385} {\bibfield  {journal} {\bibinfo  {journal} {Science}\ }\textbf {\bibinfo {volume} {373}},\ \bibinfo {pages} {1359} (\bibinfo {year} {2021})}\BibitemShut {NoStop}%
\bibitem [{\citenamefont {Helson}\ \emph {et~al.}(2023)\citenamefont {Helson}, \citenamefont {Zwettler}, \citenamefont {Mivehvar}, \citenamefont {Colella}, \citenamefont {Roux}, \citenamefont {Konishi}, \citenamefont {Ritsch},\ and\ \citenamefont {Brantut}}]{helson:2023aa}%
  \BibitemOpen
  \bibfield  {author} {\bibinfo {author} {\bibfnamefont {V.}~\bibnamefont {Helson}}, \bibinfo {author} {\bibfnamefont {T.}~\bibnamefont {Zwettler}}, \bibinfo {author} {\bibfnamefont {F.}~\bibnamefont {Mivehvar}}, \bibinfo {author} {\bibfnamefont {E.}~\bibnamefont {Colella}}, \bibinfo {author} {\bibfnamefont {K.}~\bibnamefont {Roux}}, \bibinfo {author} {\bibfnamefont {H.}~\bibnamefont {Konishi}}, \bibinfo {author} {\bibfnamefont {H.}~\bibnamefont {Ritsch}},\ and\ \bibinfo {author} {\bibfnamefont {J.-P.}\ \bibnamefont {Brantut}},\ }\bibfield  {title} {\bibinfo {title} {Density-wave ordering in a unitary fermi gas with photon-mediated interactions},\ }\href {https://doi.org/https://doi.org/10.1038/s41586-023-06018-3} {\bibfield  {journal} {\bibinfo  {journal} {Nature}\ }\textbf {\bibinfo {volume} {618}},\ \bibinfo {pages} {716} (\bibinfo {year} {2023})}\BibitemShut {NoStop}%
\bibitem [{\citenamefont {L{\'e}onard}\ \emph {et~al.}(2017)\citenamefont {L{\'e}onard}, \citenamefont {Morales}, \citenamefont {Zupancic}, \citenamefont {Donner},\ and\ \citenamefont {Esslinger}}]{leonard:2017wx}%
  \BibitemOpen
  \bibfield  {author} {\bibinfo {author} {\bibfnamefont {J.}~\bibnamefont {L{\'e}onard}}, \bibinfo {author} {\bibfnamefont {A.}~\bibnamefont {Morales}}, \bibinfo {author} {\bibfnamefont {P.}~\bibnamefont {Zupancic}}, \bibinfo {author} {\bibfnamefont {T.}~\bibnamefont {Donner}},\ and\ \bibinfo {author} {\bibfnamefont {T.}~\bibnamefont {Esslinger}},\ }\bibfield  {title} {\bibinfo {title} {Monitoring and manipulating higgs and goldstone modes in a supersolid quantum gas},\ }\href {https://doi.org/https://doi.org/10.1126/science.aan2608} {\bibfield  {journal} {\bibinfo  {journal} {Science}\ }\textbf {\bibinfo {volume} {358}},\ \bibinfo {pages} {1415} (\bibinfo {year} {2017})}\BibitemShut {NoStop}%
\bibitem [{\citenamefont {Dogra}\ \emph {et~al.}(2019)\citenamefont {Dogra}, \citenamefont {Landini}, \citenamefont {Kroeger}, \citenamefont {Hruby}, \citenamefont {Donner},\ and\ \citenamefont {Esslinger}}]{Dogra:2019aa}%
  \BibitemOpen
  \bibfield  {author} {\bibinfo {author} {\bibfnamefont {N.}~\bibnamefont {Dogra}}, \bibinfo {author} {\bibfnamefont {M.}~\bibnamefont {Landini}}, \bibinfo {author} {\bibfnamefont {K.}~\bibnamefont {Kroeger}}, \bibinfo {author} {\bibfnamefont {L.}~\bibnamefont {Hruby}}, \bibinfo {author} {\bibfnamefont {T.}~\bibnamefont {Donner}},\ and\ \bibinfo {author} {\bibfnamefont {T.}~\bibnamefont {Esslinger}},\ }\bibfield  {title} {\bibinfo {title} {Dissipation-induced structural instability and chiral dynamics in a quantum gas},\ }\href {https://doi.org/https://doi.org/10.48550/arXiv.1901.05974} {\bibfield  {journal} {\bibinfo  {journal} {Science}\ }\textbf {\bibinfo {volume} {366}},\ \bibinfo {pages} {1496} (\bibinfo {year} {2019})}\BibitemShut {NoStop}%
\bibitem [{\citenamefont {Dreon}\ \emph {et~al.}(2022)\citenamefont {Dreon}, \citenamefont {Baumg{\"a}rtner}, \citenamefont {Li}, \citenamefont {Hertlein}, \citenamefont {Esslinger},\ and\ \citenamefont {Donner}}]{dreon:2022aa}%
  \BibitemOpen
  \bibfield  {author} {\bibinfo {author} {\bibfnamefont {D.}~\bibnamefont {Dreon}}, \bibinfo {author} {\bibfnamefont {A.}~\bibnamefont {Baumg{\"a}rtner}}, \bibinfo {author} {\bibfnamefont {X.}~\bibnamefont {Li}}, \bibinfo {author} {\bibfnamefont {S.}~\bibnamefont {Hertlein}}, \bibinfo {author} {\bibfnamefont {T.}~\bibnamefont {Esslinger}},\ and\ \bibinfo {author} {\bibfnamefont {T.}~\bibnamefont {Donner}},\ }\bibfield  {title} {\bibinfo {title} {Self-oscillating pump in a topological dissipative atom--cavity system},\ }\href {https://doi.org/https://doi.org/10.1038/s41586-022-04970-0} {\bibfield  {journal} {\bibinfo  {journal} {Nature}\ }\textbf {\bibinfo {volume} {608}},\ \bibinfo {pages} {494} (\bibinfo {year} {2022})}\BibitemShut {NoStop}%
\bibitem [{\citenamefont {Wu}\ \emph {et~al.}(2023)\citenamefont {Wu}, \citenamefont {Fan}, \citenamefont {Zhang}, \citenamefont {Qi},\ and\ \citenamefont {Wu}}]{wu:2023aa}%
  \BibitemOpen
  \bibfield  {author} {\bibinfo {author} {\bibfnamefont {Z.}~\bibnamefont {Wu}}, \bibinfo {author} {\bibfnamefont {J.}~\bibnamefont {Fan}}, \bibinfo {author} {\bibfnamefont {X.}~\bibnamefont {Zhang}}, \bibinfo {author} {\bibfnamefont {J.}~\bibnamefont {Qi}},\ and\ \bibinfo {author} {\bibfnamefont {H.}~\bibnamefont {Wu}},\ }\bibfield  {title} {\bibinfo {title} {Signatures of prethermalization in a quenched cavity-mediated long-range interacting fermi gas},\ }\href {https://doi.org/https://doi.org/10.1103/PhysRevLett.131.243401} {\bibfield  {journal} {\bibinfo  {journal} {Phys. Rev. Lett.}\ }\textbf {\bibinfo {volume} {131}},\ \bibinfo {pages} {243401} (\bibinfo {year} {2023})}\BibitemShut {NoStop}%
\bibitem [{\citenamefont {Zwettler}\ \emph {et~al.}(2025{\natexlab{a}})\citenamefont {Zwettler}, \citenamefont {Del~Pace}, \citenamefont {Marijanovic}, \citenamefont {Chattopadhyay}, \citenamefont {B\"uhler}, \citenamefont {Halati}, \citenamefont {Skolc}, \citenamefont {Tolle}, \citenamefont {Helson}, \citenamefont {Bolognini}, \citenamefont {Fabre}, \citenamefont {Uchino}, \citenamefont {Giamarchi}, \citenamefont {Demler},\ and\ \citenamefont {Brantut}}]{zwettler:2025ab}%
  \BibitemOpen
  \bibfield  {author} {\bibinfo {author} {\bibfnamefont {T.}~\bibnamefont {Zwettler}}, \bibinfo {author} {\bibfnamefont {G.}~\bibnamefont {Del~Pace}}, \bibinfo {author} {\bibfnamefont {F.}~\bibnamefont {Marijanovic}}, \bibinfo {author} {\bibfnamefont {S.}~\bibnamefont {Chattopadhyay}}, \bibinfo {author} {\bibfnamefont {T.}~\bibnamefont {B\"uhler}}, \bibinfo {author} {\bibfnamefont {C.-M.}\ \bibnamefont {Halati}}, \bibinfo {author} {\bibfnamefont {L.}~\bibnamefont {Skolc}}, \bibinfo {author} {\bibfnamefont {L.}~\bibnamefont {Tolle}}, \bibinfo {author} {\bibfnamefont {V.}~\bibnamefont {Helson}}, \bibinfo {author} {\bibfnamefont {G.}~\bibnamefont {Bolognini}}, \bibinfo {author} {\bibfnamefont {A.}~\bibnamefont {Fabre}}, \bibinfo {author} {\bibfnamefont {S.}~\bibnamefont {Uchino}}, \bibinfo {author} {\bibfnamefont {T.}~\bibnamefont {Giamarchi}}, \bibinfo {author} {\bibfnamefont {E.}~\bibnamefont {Demler}},\ and\ \bibinfo {author} {\bibfnamefont {J.~P.}\ \bibnamefont {Brantut}},\ }\bibfield  {title} {\bibinfo
  {title} {Nonequilibrium dynamics of long-range interacting fermions},\ }\href {https://doi.org/https://doi.org/10.1103/PhysRevX.15.021089} {\bibfield  {journal} {\bibinfo  {journal} {Phys. Rev. X}\ }\textbf {\bibinfo {volume} {15}},\ \bibinfo {pages} {021089} (\bibinfo {year} {2025}{\natexlab{a}})}\BibitemShut {NoStop}%
\bibitem [{\citenamefont {Gerbier}\ \emph {et~al.}(2005)\citenamefont {Gerbier}, \citenamefont {Widera}, \citenamefont {F\"olling}, \citenamefont {Mandel}, \citenamefont {Gericke},\ and\ \citenamefont {Bloch}}]{gerbier:2005aa}%
  \BibitemOpen
  \bibfield  {author} {\bibinfo {author} {\bibfnamefont {F.}~\bibnamefont {Gerbier}}, \bibinfo {author} {\bibfnamefont {A.}~\bibnamefont {Widera}}, \bibinfo {author} {\bibfnamefont {S.}~\bibnamefont {F\"olling}}, \bibinfo {author} {\bibfnamefont {O.}~\bibnamefont {Mandel}}, \bibinfo {author} {\bibfnamefont {T.}~\bibnamefont {Gericke}},\ and\ \bibinfo {author} {\bibfnamefont {I.}~\bibnamefont {Bloch}},\ }\bibfield  {title} {\bibinfo {title} {Interference pattern and visibility of a mott insulator},\ }\href {https://doi.org/https://doi.org/10.1103/PhysRevA.72.053606} {\bibfield  {journal} {\bibinfo  {journal} {Phys. Rev. A}\ }\textbf {\bibinfo {volume} {72}},\ \bibinfo {pages} {053606} (\bibinfo {year} {2005})}\BibitemShut {NoStop}%
\bibitem [{\citenamefont {Gericke}\ \emph {et~al.}(2008)\citenamefont {Gericke}, \citenamefont {Wurtz}, \citenamefont {Reitz}, \citenamefont {Langen},\ and\ \citenamefont {Ott}}]{Gericke:2008aa}%
  \BibitemOpen
  \bibfield  {author} {\bibinfo {author} {\bibfnamefont {T.}~\bibnamefont {Gericke}}, \bibinfo {author} {\bibfnamefont {P.}~\bibnamefont {Wurtz}}, \bibinfo {author} {\bibfnamefont {D.}~\bibnamefont {Reitz}}, \bibinfo {author} {\bibfnamefont {T.}~\bibnamefont {Langen}},\ and\ \bibinfo {author} {\bibfnamefont {H.}~\bibnamefont {Ott}},\ }\bibfield  {title} {\bibinfo {title} {High-resolution scanning electron microscopy of an ultracold quantum gas},\ }\href {https://doi.org/https://doi.org/10.1038/nphys1102} {\bibfield  {journal} {\bibinfo  {journal} {Nat Phys}\ }\textbf {\bibinfo {volume} {4}},\ \bibinfo {pages} {949} (\bibinfo {year} {2008})}\BibitemShut {NoStop}%
\bibitem [{\citenamefont {{Gemelke}}\ \emph {et~al.}(2009)\citenamefont {{Gemelke}}, \citenamefont {{Zhang}}, \citenamefont {{Hung}},\ and\ \citenamefont {{Chin}}}]{Gemelke:2009aa}%
  \BibitemOpen
  \bibfield  {author} {\bibinfo {author} {\bibfnamefont {N.}~\bibnamefont {{Gemelke}}}, \bibinfo {author} {\bibfnamefont {X.}~\bibnamefont {{Zhang}}}, \bibinfo {author} {\bibfnamefont {C.}~\bibnamefont {{Hung}}},\ and\ \bibinfo {author} {\bibfnamefont {C.}~\bibnamefont {{Chin}}},\ }\bibfield  {title} {\bibinfo {title} {{In situ observation of incompressible Mott-insulating domains in ultracold atomic gases}},\ }\href {https://doi.org/https://doi.org/10.48550/arXiv.0904.1532} {\bibfield  {journal} {\bibinfo  {journal} {\nat}\ }\textbf {\bibinfo {volume} {460}},\ \bibinfo {pages} {995} (\bibinfo {year} {2009})}\BibitemShut {NoStop}%
\bibitem [{\citenamefont {{Bakr}}\ \emph {et~al.}(2009)\citenamefont {{Bakr}}, \citenamefont {{Gillen}}, \citenamefont {{Peng}}, \citenamefont {{F{\"o}lling}},\ and\ \citenamefont {{Greiner}}}]{Bakr:2009aa}%
  \BibitemOpen
  \bibfield  {author} {\bibinfo {author} {\bibfnamefont {W.~S.}\ \bibnamefont {{Bakr}}}, \bibinfo {author} {\bibfnamefont {J.~I.}\ \bibnamefont {{Gillen}}}, \bibinfo {author} {\bibfnamefont {A.}~\bibnamefont {{Peng}}}, \bibinfo {author} {\bibfnamefont {S.}~\bibnamefont {{F{\"o}lling}}},\ and\ \bibinfo {author} {\bibfnamefont {M.}~\bibnamefont {{Greiner}}},\ }\bibfield  {title} {\bibinfo {title} {{A quantum gas microscope for detecting single atoms in a Hubbard-regime optical lattice}},\ }\href {https://doi.org/https://doi.org/10.1038/nature08482} {\bibfield  {journal} {\bibinfo  {journal} {\nat}\ }\textbf {\bibinfo {volume} {462}},\ \bibinfo {pages} {74} (\bibinfo {year} {2009})}\BibitemShut {NoStop}%
\bibitem [{\citenamefont {{Sherson}}\ \emph {et~al.}(2010)\citenamefont {{Sherson}}, \citenamefont {{Weitenberg}}, \citenamefont {{Endres}}, \citenamefont {{Cheneau}}, \citenamefont {{Bloch}},\ and\ \citenamefont {{Kuhr}}}]{Sherson:2010aa}%
  \BibitemOpen
  \bibfield  {author} {\bibinfo {author} {\bibfnamefont {J.~F.}\ \bibnamefont {{Sherson}}}, \bibinfo {author} {\bibfnamefont {C.}~\bibnamefont {{Weitenberg}}}, \bibinfo {author} {\bibfnamefont {M.}~\bibnamefont {{Endres}}}, \bibinfo {author} {\bibfnamefont {M.}~\bibnamefont {{Cheneau}}}, \bibinfo {author} {\bibfnamefont {I.}~\bibnamefont {{Bloch}}},\ and\ \bibinfo {author} {\bibfnamefont {S.}~\bibnamefont {{Kuhr}}},\ }\bibfield  {title} {\bibinfo {title} {{Single-atom-resolved fluorescence imaging of an atomic Mott insulator}},\ }\href {https://doi.org/https://doi.org/10.1038/nature09378} {\bibfield  {journal} {\bibinfo  {journal} {\nat}\ }\textbf {\bibinfo {volume} {467}},\ \bibinfo {pages} {68} (\bibinfo {year} {2010})}\BibitemShut {NoStop}%
\bibitem [{\citenamefont {Endres}\ \emph {et~al.}(2011)\citenamefont {Endres}, \citenamefont {Cheneau}, \citenamefont {Fukuhara}, \citenamefont {Weitenberg}, \citenamefont {Schau{\ss}}, \citenamefont {Gross}, \citenamefont {Mazza}, \citenamefont {Ba{\~n}uls}, \citenamefont {Pollet}, \citenamefont {Bloch},\ and\ \citenamefont {Kuhr}}]{Endres:2011aa}%
  \BibitemOpen
  \bibfield  {author} {\bibinfo {author} {\bibfnamefont {M.}~\bibnamefont {Endres}}, \bibinfo {author} {\bibfnamefont {M.}~\bibnamefont {Cheneau}}, \bibinfo {author} {\bibfnamefont {T.}~\bibnamefont {Fukuhara}}, \bibinfo {author} {\bibfnamefont {C.}~\bibnamefont {Weitenberg}}, \bibinfo {author} {\bibfnamefont {P.}~\bibnamefont {Schau{\ss}}}, \bibinfo {author} {\bibfnamefont {C.}~\bibnamefont {Gross}}, \bibinfo {author} {\bibfnamefont {L.}~\bibnamefont {Mazza}}, \bibinfo {author} {\bibfnamefont {M.~C.}\ \bibnamefont {Ba{\~n}uls}}, \bibinfo {author} {\bibfnamefont {L.}~\bibnamefont {Pollet}}, \bibinfo {author} {\bibfnamefont {I.}~\bibnamefont {Bloch}},\ and\ \bibinfo {author} {\bibfnamefont {S.}~\bibnamefont {Kuhr}},\ }\bibfield  {title} {\bibinfo {title} {Observation of correlated particle-hole pairs and string order in low-dimensional mott insulators},\ }\href {https://doi.org/10.1126/science.1209284} {\bibfield  {journal} {\bibinfo  {journal} {Science}\ }\textbf {\bibinfo {volume} {334}},\ \bibinfo {pages}
  {200} (\bibinfo {year} {2011})}\BibitemShut {NoStop}%
\bibitem [{\citenamefont {Hilker}\ \emph {et~al.}(2017)\citenamefont {Hilker}, \citenamefont {Salomon}, \citenamefont {Grusdt}, \citenamefont {Omran}, \citenamefont {Boll}, \citenamefont {Demler}, \citenamefont {Bloch},\ and\ \citenamefont {Gross}}]{hilker:2017ab}%
  \BibitemOpen
  \bibfield  {author} {\bibinfo {author} {\bibfnamefont {T.~A.}\ \bibnamefont {Hilker}}, \bibinfo {author} {\bibfnamefont {G.}~\bibnamefont {Salomon}}, \bibinfo {author} {\bibfnamefont {F.}~\bibnamefont {Grusdt}}, \bibinfo {author} {\bibfnamefont {A.}~\bibnamefont {Omran}}, \bibinfo {author} {\bibfnamefont {M.}~\bibnamefont {Boll}}, \bibinfo {author} {\bibfnamefont {E.}~\bibnamefont {Demler}}, \bibinfo {author} {\bibfnamefont {I.}~\bibnamefont {Bloch}},\ and\ \bibinfo {author} {\bibfnamefont {C.}~\bibnamefont {Gross}},\ }\bibfield  {title} {\bibinfo {title} {Revealing hidden antiferromagnetic correlations in doped hubbard chains via string correlators},\ }\href {https://doi.org/DOI: 10.1126/science.aam8990} {\bibfield  {journal} {\bibinfo  {journal} {Science}\ }\textbf {\bibinfo {volume} {357}},\ \bibinfo {pages} {484} (\bibinfo {year} {2017})}\BibitemShut {NoStop}%
\bibitem [{\citenamefont {Preiss}\ \emph {et~al.}(2015)\citenamefont {Preiss}, \citenamefont {Ma}, \citenamefont {Tai}, \citenamefont {Lukin}, \citenamefont {Rispoli}, \citenamefont {Zupancic}, \citenamefont {Lahini}, \citenamefont {Islam},\ and\ \citenamefont {Greiner}}]{Preiss:2015ab}%
  \BibitemOpen
  \bibfield  {author} {\bibinfo {author} {\bibfnamefont {P.~M.}\ \bibnamefont {Preiss}}, \bibinfo {author} {\bibfnamefont {R.}~\bibnamefont {Ma}}, \bibinfo {author} {\bibfnamefont {M.~E.}\ \bibnamefont {Tai}}, \bibinfo {author} {\bibfnamefont {A.}~\bibnamefont {Lukin}}, \bibinfo {author} {\bibfnamefont {M.}~\bibnamefont {Rispoli}}, \bibinfo {author} {\bibfnamefont {P.}~\bibnamefont {Zupancic}}, \bibinfo {author} {\bibfnamefont {Y.}~\bibnamefont {Lahini}}, \bibinfo {author} {\bibfnamefont {R.}~\bibnamefont {Islam}},\ and\ \bibinfo {author} {\bibfnamefont {M.}~\bibnamefont {Greiner}},\ }\bibfield  {title} {\bibinfo {title} {Strongly correlated quantum walks in optical lattices},\ }\href {https://doi.org/https://doi.org/10.1126/science.1260364} {\bibfield  {journal} {\bibinfo  {journal} {Science}\ }\textbf {\bibinfo {volume} {347}},\ \bibinfo {pages} {1229} (\bibinfo {year} {2015})}\BibitemShut {NoStop}%
\bibitem [{\citenamefont {Ðorđevi{\'c}}\ \emph {et~al.}(2021)\citenamefont {Ðorđevi{\'c}}, \citenamefont {Samutpraphoot}, \citenamefont {Ocola}, \citenamefont {Bernien}, \citenamefont {Grinkemeyer}, \citenamefont {Dimitrova}, \citenamefont {Vuleti{\'c}},\ and\ \citenamefont {Lukin}}]{dordevic:2021aa}%
  \BibitemOpen
  \bibfield  {author} {\bibinfo {author} {\bibfnamefont {T.}~\bibnamefont {Ðorđevi{\'c}}}, \bibinfo {author} {\bibfnamefont {P.}~\bibnamefont {Samutpraphoot}}, \bibinfo {author} {\bibfnamefont {P.~L.}\ \bibnamefont {Ocola}}, \bibinfo {author} {\bibfnamefont {H.}~\bibnamefont {Bernien}}, \bibinfo {author} {\bibfnamefont {B.}~\bibnamefont {Grinkemeyer}}, \bibinfo {author} {\bibfnamefont {I.}~\bibnamefont {Dimitrova}}, \bibinfo {author} {\bibfnamefont {V.}~\bibnamefont {Vuleti{\'c}}},\ and\ \bibinfo {author} {\bibfnamefont {M.~D.}\ \bibnamefont {Lukin}},\ }\bibfield  {title} {\bibinfo {title} {Entanglement transport and a nanophotonic interface for atoms in optical tweezers},\ }\href {https://doi.org/10.1126/science.abi9917} {\bibfield  {journal} {\bibinfo  {journal} {Science}\ }\textbf {\bibinfo {volume} {373}},\ \bibinfo {pages} {1511} (\bibinfo {year} {2021})}\BibitemShut {NoStop}%
\bibitem [{\citenamefont {Uru\~nuela}\ \emph {et~al.}(2022)\citenamefont {Uru\~nuela}, \citenamefont {Ammenwerth}, \citenamefont {Malik}, \citenamefont {Ahlheit}, \citenamefont {Pfeifer}, \citenamefont {Alt},\ and\ \citenamefont {Meschede}}]{urunuela:2022aa}%
  \BibitemOpen
  \bibfield  {author} {\bibinfo {author} {\bibfnamefont {E.}~\bibnamefont {Uru\~nuela}}, \bibinfo {author} {\bibfnamefont {M.}~\bibnamefont {Ammenwerth}}, \bibinfo {author} {\bibfnamefont {P.}~\bibnamefont {Malik}}, \bibinfo {author} {\bibfnamefont {L.}~\bibnamefont {Ahlheit}}, \bibinfo {author} {\bibfnamefont {H.}~\bibnamefont {Pfeifer}}, \bibinfo {author} {\bibfnamefont {W.}~\bibnamefont {Alt}},\ and\ \bibinfo {author} {\bibfnamefont {D.}~\bibnamefont {Meschede}},\ }\bibfield  {title} {\bibinfo {title} {Raman imaging of atoms inside a high-bandwidth cavity},\ }\href {https://doi.org/https://doi.org/10.1103/PhysRevA.105.043321} {\bibfield  {journal} {\bibinfo  {journal} {Phys. Rev. A}\ }\textbf {\bibinfo {volume} {105}},\ \bibinfo {pages} {043321} (\bibinfo {year} {2022})}\BibitemShut {NoStop}%
\bibitem [{\citenamefont {Deist}\ \emph {et~al.}(2022)\citenamefont {Deist}, \citenamefont {Gerber}, \citenamefont {Lu}, \citenamefont {Zeiher},\ and\ \citenamefont {Stamper-Kurn}}]{deist:2022aa}%
  \BibitemOpen
  \bibfield  {author} {\bibinfo {author} {\bibfnamefont {E.}~\bibnamefont {Deist}}, \bibinfo {author} {\bibfnamefont {J.~A.}\ \bibnamefont {Gerber}}, \bibinfo {author} {\bibfnamefont {Y.-H.}\ \bibnamefont {Lu}}, \bibinfo {author} {\bibfnamefont {J.}~\bibnamefont {Zeiher}},\ and\ \bibinfo {author} {\bibfnamefont {D.~M.}\ \bibnamefont {Stamper-Kurn}},\ }\bibfield  {title} {\bibinfo {title} {Superresolution microscopy of optical fields using tweezer-trapped single atoms},\ }\href {https://doi.org/https://doi.org/10.1103/PhysRevLett.128.083201} {\bibfield  {journal} {\bibinfo  {journal} {Phys. Rev. Lett.}\ }\textbf {\bibinfo {volume} {128}},\ \bibinfo {pages} {083201} (\bibinfo {year} {2022})}\BibitemShut {NoStop}%
\bibitem [{\citenamefont {Liu}\ \emph {et~al.}(2023)\citenamefont {Liu}, \citenamefont {Wang}, \citenamefont {Yang}, \citenamefont {Wang}, \citenamefont {Fan}, \citenamefont {Guan}, \citenamefont {Li}, \citenamefont {Zhang},\ and\ \citenamefont {Zhang}}]{liu:2023aa}%
  \BibitemOpen
  \bibfield  {author} {\bibinfo {author} {\bibfnamefont {Y.}~\bibnamefont {Liu}}, \bibinfo {author} {\bibfnamefont {Z.}~\bibnamefont {Wang}}, \bibinfo {author} {\bibfnamefont {P.}~\bibnamefont {Yang}}, \bibinfo {author} {\bibfnamefont {Q.}~\bibnamefont {Wang}}, \bibinfo {author} {\bibfnamefont {Q.}~\bibnamefont {Fan}}, \bibinfo {author} {\bibfnamefont {S.}~\bibnamefont {Guan}}, \bibinfo {author} {\bibfnamefont {G.}~\bibnamefont {Li}}, \bibinfo {author} {\bibfnamefont {P.}~\bibnamefont {Zhang}},\ and\ \bibinfo {author} {\bibfnamefont {T.}~\bibnamefont {Zhang}},\ }\bibfield  {title} {\bibinfo {title} {Realization of strong coupling between deterministic single-atom arrays and a high-finesse miniature optical cavity},\ }\href {https://doi.org/https://doi.org/10.1103/PhysRevLett.130.173601} {\bibfield  {journal} {\bibinfo  {journal} {Phys. Rev. Lett.}\ }\textbf {\bibinfo {volume} {130}},\ \bibinfo {pages} {173601} (\bibinfo {year} {2023})}\BibitemShut {NoStop}%
\bibitem [{\citenamefont {Zhang}\ \emph {et~al.}(2024)\citenamefont {Zhang}, \citenamefont {Yu}, \citenamefont {Zhang}, \citenamefont {Xiang},\ and\ \citenamefont {Zhang}}]{zhang:2024aa}%
  \BibitemOpen
  \bibfield  {author} {\bibinfo {author} {\bibfnamefont {X.}~\bibnamefont {Zhang}}, \bibinfo {author} {\bibfnamefont {Z.}~\bibnamefont {Yu}}, \bibinfo {author} {\bibfnamefont {H.}~\bibnamefont {Zhang}}, \bibinfo {author} {\bibfnamefont {D.}~\bibnamefont {Xiang}},\ and\ \bibinfo {author} {\bibfnamefont {H.}~\bibnamefont {Zhang}},\ }\bibfield  {title} {\bibinfo {title} {Cavity dark mode mediated by atom array without atomic scattering loss},\ }\href {https://doi.org/https://doi.org/10.1103/PhysRevResearch.6.L042026} {\bibfield  {journal} {\bibinfo  {journal} {Phys. Rev. Res.}\ }\textbf {\bibinfo {volume} {6}},\ \bibinfo {pages} {L042026} (\bibinfo {year} {2024})}\BibitemShut {NoStop}%
\bibitem [{\citenamefont {Hartung}\ \emph {et~al.}(2024)\citenamefont {Hartung}, \citenamefont {Seubert}, \citenamefont {Welte}, \citenamefont {Distante},\ and\ \citenamefont {Rempe}}]{hartung:2024aa}%
  \BibitemOpen
  \bibfield  {author} {\bibinfo {author} {\bibfnamefont {L.}~\bibnamefont {Hartung}}, \bibinfo {author} {\bibfnamefont {M.}~\bibnamefont {Seubert}}, \bibinfo {author} {\bibfnamefont {S.}~\bibnamefont {Welte}}, \bibinfo {author} {\bibfnamefont {E.}~\bibnamefont {Distante}},\ and\ \bibinfo {author} {\bibfnamefont {G.}~\bibnamefont {Rempe}},\ }\bibfield  {title} {\bibinfo {title} {A quantum-network register assembled with optical tweezers in an optical cavity},\ }\href {https://doi.org/https://doi.org/10.1126/science.ado6471} {\bibfield  {journal} {\bibinfo  {journal} {Science}\ }\textbf {\bibinfo {volume} {385}},\ \bibinfo {pages} {179} (\bibinfo {year} {2024})}\BibitemShut {NoStop}%
\bibitem [{\citenamefont {Hu}\ \emph {et~al.}(2025)\citenamefont {Hu}, \citenamefont {Sinclair}, \citenamefont {Bytyqi}, \citenamefont {Chong}, \citenamefont {Rudelis}, \citenamefont {Ramette}, \citenamefont {Vendeiro},\ and\ \citenamefont {Vuleti\ifmmode~\acute{c}\else \'{c}\fi{}}}]{hu:2025aa}%
  \BibitemOpen
  \bibfield  {author} {\bibinfo {author} {\bibfnamefont {B.}~\bibnamefont {Hu}}, \bibinfo {author} {\bibfnamefont {J.}~\bibnamefont {Sinclair}}, \bibinfo {author} {\bibfnamefont {E.}~\bibnamefont {Bytyqi}}, \bibinfo {author} {\bibfnamefont {M.}~\bibnamefont {Chong}}, \bibinfo {author} {\bibfnamefont {A.}~\bibnamefont {Rudelis}}, \bibinfo {author} {\bibfnamefont {J.}~\bibnamefont {Ramette}}, \bibinfo {author} {\bibfnamefont {Z.}~\bibnamefont {Vendeiro}},\ and\ \bibinfo {author} {\bibfnamefont {V.}~\bibnamefont {Vuleti\ifmmode~\acute{c}\else \'{c}\fi{}}},\ }\bibfield  {title} {\bibinfo {title} {Site-selective cavity readout and classical error correction of a 5-bit atomic register},\ }\href {https://doi.org/https://doi.org/10.48550/arXiv.2408.15329} {\bibfield  {journal} {\bibinfo  {journal} {Phys. Rev. Lett.}\ }\textbf {\bibinfo {volume} {134}},\ \bibinfo {pages} {120801} (\bibinfo {year} {2025})}\BibitemShut {NoStop}%
\bibitem [{\citenamefont {Orsi}\ \emph {et~al.}(2024)\citenamefont {Orsi}, \citenamefont {Sauerwein}, \citenamefont {Bhatt}, \citenamefont {Faltinath}, \citenamefont {Fedotova}, \citenamefont {Reiter}, \citenamefont {Cantat-Moltrecht},\ and\ \citenamefont {Brantut}}]{orsi:2024ab}%
  \BibitemOpen
  \bibfield  {author} {\bibinfo {author} {\bibfnamefont {F.}~\bibnamefont {Orsi}}, \bibinfo {author} {\bibfnamefont {N.}~\bibnamefont {Sauerwein}}, \bibinfo {author} {\bibfnamefont {R.~P.}\ \bibnamefont {Bhatt}}, \bibinfo {author} {\bibfnamefont {J.}~\bibnamefont {Faltinath}}, \bibinfo {author} {\bibfnamefont {E.}~\bibnamefont {Fedotova}}, \bibinfo {author} {\bibfnamefont {N.}~\bibnamefont {Reiter}}, \bibinfo {author} {\bibfnamefont {T.}~\bibnamefont {Cantat-Moltrecht}},\ and\ \bibinfo {author} {\bibfnamefont {J.-P.}\ \bibnamefont {Brantut}},\ }\bibfield  {title} {\bibinfo {title} {Cavity microscope for micrometer-scale control of atom-photon interactions},\ }\href {https://doi.org/https://doi.org/10.1103/PRXQuantum.5.040333} {\bibfield  {journal} {\bibinfo  {journal} {PRX Quantum}\ }\textbf {\bibinfo {volume} {5}},\ \bibinfo {pages} {040333} (\bibinfo {year} {2024})}\BibitemShut {NoStop}%
\bibitem [{\citenamefont {Zwettler}\ \emph {et~al.}(2025{\natexlab{b}})\citenamefont {Zwettler}, \citenamefont {Marijanovic}, \citenamefont {B{\"u}hler}, \citenamefont {Chattopadhyay}, \citenamefont {Del~Pace}, \citenamefont {Skolc}, \citenamefont {Helson}, \citenamefont {Uchino}, \citenamefont {Demler},\ and\ \citenamefont {Brantut}}]{zwettler:2025ac}%
  \BibitemOpen
  \bibfield  {author} {\bibinfo {author} {\bibfnamefont {T.}~\bibnamefont {Zwettler}}, \bibinfo {author} {\bibfnamefont {F.}~\bibnamefont {Marijanovic}}, \bibinfo {author} {\bibfnamefont {T.}~\bibnamefont {B{\"u}hler}}, \bibinfo {author} {\bibfnamefont {S.}~\bibnamefont {Chattopadhyay}}, \bibinfo {author} {\bibfnamefont {G.}~\bibnamefont {Del~Pace}}, \bibinfo {author} {\bibfnamefont {L.}~\bibnamefont {Skolc}}, \bibinfo {author} {\bibfnamefont {V.}~\bibnamefont {Helson}}, \bibinfo {author} {\bibfnamefont {S.}~\bibnamefont {Uchino}}, \bibinfo {author} {\bibfnamefont {E.}~\bibnamefont {Demler}},\ and\ \bibinfo {author} {\bibfnamefont {J.-P.}\ \bibnamefont {Brantut}},\ }\bibfield  {title} {\bibinfo {title} {Cavity-mediated charge and pair-density waves in a unitary {{Fermi}} gas},\ }\href {https://doi.org/10.1038/s41467-025-67184-8} {\bibfield  {journal} {\bibinfo  {journal} {Nature Communications}\ }\textbf {\bibinfo {volume} {17}},\ \bibinfo {pages} {496} (\bibinfo {year} {2025}{\natexlab{b}})}\BibitemShut
  {NoStop}%
\bibitem [{\citenamefont {Roux}\ \emph {et~al.}(2020)\citenamefont {Roux}, \citenamefont {Konishi}, \citenamefont {Helson},\ and\ \citenamefont {Brantut}}]{Roux:2020aa}%
  \BibitemOpen
  \bibfield  {author} {\bibinfo {author} {\bibfnamefont {K.}~\bibnamefont {Roux}}, \bibinfo {author} {\bibfnamefont {H.}~\bibnamefont {Konishi}}, \bibinfo {author} {\bibfnamefont {V.}~\bibnamefont {Helson}},\ and\ \bibinfo {author} {\bibfnamefont {J.-P.}\ \bibnamefont {Brantut}},\ }\bibfield  {title} {\bibinfo {title} {Strongly correlated fermions strongly coupled to light},\ }\href {https://doi.org/https://doi.org/10.1038/s41467-020-16767-8} {\bibfield  {journal} {\bibinfo  {journal} {Nature Communications}\ }\textbf {\bibinfo {volume} {11}},\ \bibinfo {pages} {2974} (\bibinfo {year} {2020})}\BibitemShut {NoStop}%
\bibitem [{\citenamefont {Bühler}\ \emph {et~al.}(2025)\citenamefont {Bühler}, \citenamefont {Zwettler}, \citenamefont {Bolognini}, \citenamefont {Fabre}, \citenamefont {Helson}, \citenamefont {{Del Pace}},\ and\ \citenamefont {Brantut}}]{Buhler:2025}%
  \BibitemOpen
  \bibfield  {author} {\bibinfo {author} {\bibfnamefont {T.}~\bibnamefont {Bühler}}, \bibinfo {author} {\bibfnamefont {T.}~\bibnamefont {Zwettler}}, \bibinfo {author} {\bibfnamefont {G.}~\bibnamefont {Bolognini}}, \bibinfo {author} {\bibfnamefont {A.}~\bibnamefont {Fabre}}, \bibinfo {author} {\bibfnamefont {V.}~\bibnamefont {Helson}}, \bibinfo {author} {\bibfnamefont {G.}~\bibnamefont {{Del Pace}}},\ and\ \bibinfo {author} {\bibfnamefont {J.-P.}\ \bibnamefont {Brantut}},\ }\bibfield  {title} {\bibinfo {title} {{Direct production of fermionic superfluids in a cavity-enhanced optical dipole trap}},\ }\href {https://doi.org/10.21468/SciPostPhys.18.4.133} {\bibfield  {journal} {\bibinfo  {journal} {SciPost Phys.}\ }\textbf {\bibinfo {volume} {18}},\ \bibinfo {pages} {133} (\bibinfo {year} {2025})}\BibitemShut {NoStop}%
\bibitem [{sup()}]{supp}%
  \BibitemOpen
  \href@noop {} {}\bibinfo {note} {See Supplemental Material at [URL] for details on the experimental procedures, theoretical descriptions and supplementary data, which includes Refs. [48-55].}\BibitemShut {Stop}%
\bibitem [{\citenamefont {{Reinaudi}}\ \emph {et~al.}(2007)\citenamefont {{Reinaudi}}, \citenamefont {{Lahaye}}, \citenamefont {{Wang}},\ and\ \citenamefont {{Gu{\'e}ry-Odelin}}}]{Reinaudi:2007aa}%
  \BibitemOpen
  \bibfield  {author} {\bibinfo {author} {\bibfnamefont {G.}~\bibnamefont {{Reinaudi}}}, \bibinfo {author} {\bibfnamefont {T.}~\bibnamefont {{Lahaye}}}, \bibinfo {author} {\bibfnamefont {Z.}~\bibnamefont {{Wang}}},\ and\ \bibinfo {author} {\bibfnamefont {D.}~\bibnamefont {{Gu{\'e}ry-Odelin}}},\ }\bibfield  {title} {\bibinfo {title} {{Strong saturation absorption imaging of dense clouds of ultracold atoms}},\ }\href {https://doi.org/https://doi.org/10.1364/OL.32.003143} {\bibfield  {journal} {\bibinfo  {journal} {Optics Letters}\ }\textbf {\bibinfo {volume} {32}},\ \bibinfo {pages} {3143} (\bibinfo {year} {2007})}\BibitemShut {NoStop}%
\bibitem [{\citenamefont {Horikoshi}\ \emph {et~al.}(2017)\citenamefont {Horikoshi}, \citenamefont {Ito}, \citenamefont {Ikemachi}, \citenamefont {Aratake}, \citenamefont {Kuwata-Gonokami},\ and\ \citenamefont {Koashi}}]{Horikoshi:2017ab}%
  \BibitemOpen
  \bibfield  {author} {\bibinfo {author} {\bibfnamefont {M.}~\bibnamefont {Horikoshi}}, \bibinfo {author} {\bibfnamefont {A.}~\bibnamefont {Ito}}, \bibinfo {author} {\bibfnamefont {T.}~\bibnamefont {Ikemachi}}, \bibinfo {author} {\bibfnamefont {Y.}~\bibnamefont {Aratake}}, \bibinfo {author} {\bibfnamefont {M.}~\bibnamefont {Kuwata-Gonokami}},\ and\ \bibinfo {author} {\bibfnamefont {M.}~\bibnamefont {Koashi}},\ }\bibfield  {title} {\bibinfo {title} {Appropriate probe condition for absorption imaging of ultracold 6li atoms},\ }\href {https://doi.org/https://doi.org/10.48550/arXiv.1608.07152} {\bibfield  {journal} {\bibinfo  {journal} {Journal of the Physical Society of Japan}\ }\textbf {\bibinfo {volume} {86}},\ \bibinfo {pages} {104301} (\bibinfo {year} {2017})}\BibitemShut {NoStop}%
\bibitem [{\citenamefont {Hueck}\ \emph {et~al.}(2017)\citenamefont {Hueck}, \citenamefont {Luick}, \citenamefont {Sobirey}, \citenamefont {Siegl}, \citenamefont {Lompe}, \citenamefont {Moritz}, \citenamefont {Clark},\ and\ \citenamefont {Chin}}]{hueck_calibrating_2017}%
  \BibitemOpen
  \bibfield  {author} {\bibinfo {author} {\bibfnamefont {K.}~\bibnamefont {Hueck}}, \bibinfo {author} {\bibfnamefont {N.}~\bibnamefont {Luick}}, \bibinfo {author} {\bibfnamefont {L.}~\bibnamefont {Sobirey}}, \bibinfo {author} {\bibfnamefont {J.}~\bibnamefont {Siegl}}, \bibinfo {author} {\bibfnamefont {T.}~\bibnamefont {Lompe}}, \bibinfo {author} {\bibfnamefont {H.}~\bibnamefont {Moritz}}, \bibinfo {author} {\bibfnamefont {L.~W.}\ \bibnamefont {Clark}},\ and\ \bibinfo {author} {\bibfnamefont {C.}~\bibnamefont {Chin}},\ }\bibfield  {title} {\bibinfo {title} {Calibrating {High} {Intensity} {Absorption} {Imaging} of {Ultracold} {Atoms}},\ }\href {https://doi.org/https://doi.org/10.48550/arXiv.1702.01943} {\bibfield  {journal} {\bibinfo  {journal} {Optics Express}\ }\textbf {\bibinfo {volume} {25}},\ \bibinfo {pages} {8670} (\bibinfo {year} {2017})}\BibitemShut {NoStop}%
\bibitem [{\citenamefont {Marijanovi{\'c}}\ \emph {et~al.}(2026)\citenamefont {Marijanovi{\'c}}, \citenamefont {Chattopadhyay}, \citenamefont {Skolc}, \citenamefont {Zwettler}, \citenamefont {Halati}, \citenamefont {J{\"a}ger}, \citenamefont {Giamarchi}, \citenamefont {Brantut},\ and\ \citenamefont {Demler}}]{marijanovic:2024aa}%
  \BibitemOpen
  \bibfield  {author} {\bibinfo {author} {\bibfnamefont {F.}~\bibnamefont {Marijanovi{\'c}}}, \bibinfo {author} {\bibfnamefont {S.}~\bibnamefont {Chattopadhyay}}, \bibinfo {author} {\bibfnamefont {L.}~\bibnamefont {Skolc}}, \bibinfo {author} {\bibfnamefont {T.}~\bibnamefont {Zwettler}}, \bibinfo {author} {\bibfnamefont {C.-M.}\ \bibnamefont {Halati}}, \bibinfo {author} {\bibfnamefont {S.~B.}\ \bibnamefont {J{\"a}ger}}, \bibinfo {author} {\bibfnamefont {T.}~\bibnamefont {Giamarchi}}, \bibinfo {author} {\bibfnamefont {J.-P.}\ \bibnamefont {Brantut}},\ and\ \bibinfo {author} {\bibfnamefont {E.}~\bibnamefont {Demler}},\ }\bibfield  {title} {\bibinfo {title} {Quench instabilities of a strongly interacting quantum gas in an optical cavity},\ }\href {https://doi.org/https://doi.org/10.1103/w2vx-t1mr} {\bibfield  {journal} {\bibinfo  {journal} {Phys. Rev. Lett.}\ }\textbf {\bibinfo {volume} {136}},\ \bibinfo {pages} {193401} (\bibinfo {year} {2026})}\BibitemShut {NoStop}%
\bibitem [{\citenamefont {Sethna}(2021)}]{sethna2021statistical}%
  \BibitemOpen
  \bibfield  {author} {\bibinfo {author} {\bibfnamefont {J.~P.}\ \bibnamefont {Sethna}},\ }\href@noop {} {\emph {\bibinfo {title} {Statistical mechanics: entropy, order parameters, and complexity}}},\ Vol.~\bibinfo {volume} {14}\ (\bibinfo  {publisher} {Oxford University Press},\ \bibinfo {year} {2021})\BibitemShut {NoStop}%
\bibitem [{\citenamefont {Mivehvar}\ \emph {et~al.}(2021)\citenamefont {Mivehvar}, \citenamefont {Piazza}, \citenamefont {Donner},\ and\ \citenamefont {Ritsch}}]{mivehvar:2021aa}%
  \BibitemOpen
  \bibfield  {author} {\bibinfo {author} {\bibfnamefont {F.}~\bibnamefont {Mivehvar}}, \bibinfo {author} {\bibfnamefont {F.}~\bibnamefont {Piazza}}, \bibinfo {author} {\bibfnamefont {T.}~\bibnamefont {Donner}},\ and\ \bibinfo {author} {\bibfnamefont {H.}~\bibnamefont {Ritsch}},\ }\bibfield  {title} {\bibinfo {title} {Cavity qed with quantum gases: new paradigms in many-body physics},\ }\href {https://doi.org/https://doi.org/10.48550/arXiv.2102.04473} {\bibfield  {journal} {\bibinfo  {journal} {Advances in Physics}\ }\textbf {\bibinfo {volume} {70}},\ \bibinfo {pages} {1} (\bibinfo {year} {2021})}\BibitemShut {NoStop}%
\bibitem [{\citenamefont {Halati}\ \emph {et~al.}(2020{\natexlab{a}})\citenamefont {Halati}, \citenamefont {Sheikhan}, \citenamefont {Ritsch},\ and\ \citenamefont {Kollath}}]{halati:2020aa}%
  \BibitemOpen
  \bibfield  {author} {\bibinfo {author} {\bibfnamefont {C.-M.}\ \bibnamefont {Halati}}, \bibinfo {author} {\bibfnamefont {A.}~\bibnamefont {Sheikhan}}, \bibinfo {author} {\bibfnamefont {H.}~\bibnamefont {Ritsch}},\ and\ \bibinfo {author} {\bibfnamefont {C.}~\bibnamefont {Kollath}},\ }\bibfield  {title} {\bibinfo {title} {Numerically exact treatment of many-body self-organization in a cavity},\ }\href {https://doi.org/https://doi.org/10.1103/PhysRevLett.125.093604} {\bibfield  {journal} {\bibinfo  {journal} {Phys. Rev. Lett.}\ }\textbf {\bibinfo {volume} {125}},\ \bibinfo {pages} {093604} (\bibinfo {year} {2020}{\natexlab{a}})}\BibitemShut {NoStop}%
\bibitem [{\citenamefont {Halati}\ \emph {et~al.}(2020{\natexlab{b}})\citenamefont {Halati}, \citenamefont {Sheikhan},\ and\ \citenamefont {Kollath}}]{halati:2020ab}%
  \BibitemOpen
  \bibfield  {author} {\bibinfo {author} {\bibfnamefont {C.-M.}\ \bibnamefont {Halati}}, \bibinfo {author} {\bibfnamefont {A.}~\bibnamefont {Sheikhan}},\ and\ \bibinfo {author} {\bibfnamefont {C.}~\bibnamefont {Kollath}},\ }\bibfield  {title} {\bibinfo {title} {Theoretical methods to treat a single dissipative bosonic mode coupled globally to an interacting many-body system},\ }\href {https://doi.org/https://doi.org/10.48550/arXiv.2004.11807} {\bibfield  {journal} {\bibinfo  {journal} {Phys. Rev. Res.}\ }\textbf {\bibinfo {volume} {2}},\ \bibinfo {pages} {043255} (\bibinfo {year} {2020}{\natexlab{b}})}\BibitemShut {NoStop}%
\bibitem [{\citenamefont {Partridge}\ \emph {et~al.}(2006)\citenamefont {Partridge}, \citenamefont {Li}, \citenamefont {Kamar}, \citenamefont {Liao},\ and\ \citenamefont {Hulet}}]{Partridge:2006fk}%
  \BibitemOpen
  \bibfield  {author} {\bibinfo {author} {\bibfnamefont {G.~B.}\ \bibnamefont {Partridge}}, \bibinfo {author} {\bibfnamefont {W.}~\bibnamefont {Li}}, \bibinfo {author} {\bibfnamefont {R.~I.}\ \bibnamefont {Kamar}}, \bibinfo {author} {\bibfnamefont {Y.-a.}\ \bibnamefont {Liao}},\ and\ \bibinfo {author} {\bibfnamefont {R.~G.}\ \bibnamefont {Hulet}},\ }\bibfield  {title} {\bibinfo {title} {Pairing and phase separation in a polarized fermi gas},\ }\href {https://doi.org/DOI: 10.1126/science.1122876} {\bibfield  {journal} {\bibinfo  {journal} {Science}\ }\textbf {\bibinfo {volume} {311}},\ \bibinfo {pages} {503} (\bibinfo {year} {2006})}\BibitemShut {NoStop}%
\bibitem [{\citenamefont {Shin}\ \emph {et~al.}(2006)\citenamefont {Shin}, \citenamefont {Zwierlein}, \citenamefont {Schunck}, \citenamefont {Schirotzek},\ and\ \citenamefont {Ketterle}}]{Shin:2006aa}%
  \BibitemOpen
  \bibfield  {author} {\bibinfo {author} {\bibfnamefont {Y.}~\bibnamefont {Shin}}, \bibinfo {author} {\bibfnamefont {M.~W.}\ \bibnamefont {Zwierlein}}, \bibinfo {author} {\bibfnamefont {C.~H.}\ \bibnamefont {Schunck}}, \bibinfo {author} {\bibfnamefont {A.}~\bibnamefont {Schirotzek}},\ and\ \bibinfo {author} {\bibfnamefont {W.}~\bibnamefont {Ketterle}},\ }\bibfield  {title} {\bibinfo {title} {Observation of phase separation in a strongly interacting imbalanced fermi gas},\ }\href {https://doi.org/https://doi.org/10.1103/PhysRevLett.97.030401} {\bibfield  {journal} {\bibinfo  {journal} {Phys. Rev. Lett.}\ }\textbf {\bibinfo {volume} {97}},\ \bibinfo {pages} {030401} (\bibinfo {year} {2006})}\BibitemShut {NoStop}%
\bibitem [{\citenamefont {Shin}\ \emph {et~al.}(2007)\citenamefont {Shin}, \citenamefont {Schunck}, \citenamefont {Schirotzek},\ and\ \citenamefont {Ketterle}}]{Shin:2007aa}%
  \BibitemOpen
  \bibfield  {author} {\bibinfo {author} {\bibfnamefont {Y.}~\bibnamefont {Shin}}, \bibinfo {author} {\bibfnamefont {C.~H.}\ \bibnamefont {Schunck}}, \bibinfo {author} {\bibfnamefont {A.}~\bibnamefont {Schirotzek}},\ and\ \bibinfo {author} {\bibfnamefont {W.}~\bibnamefont {Ketterle}},\ }\bibfield  {title} {\bibinfo {title} {Tomographic rf spectroscopy of a trapped fermi gas at unitarity},\ }\href {https://doi.org/https://doi.org/10.1103/PhysRevLett.99.090403} {\bibfield  {journal} {\bibinfo  {journal} {Phys. Rev. Lett.}\ }\textbf {\bibinfo {volume} {99}},\ \bibinfo {pages} {090403} (\bibinfo {year} {2007})}\BibitemShut {NoStop}%
\bibitem [{\citenamefont {Bonifacio}\ \emph {et~al.}(2024)\citenamefont {Bonifacio}, \citenamefont {Piazza},\ and\ \citenamefont {Donner}}]{bonifacio:2024aa}%
  \BibitemOpen
  \bibfield  {author} {\bibinfo {author} {\bibfnamefont {M.}~\bibnamefont {Bonifacio}}, \bibinfo {author} {\bibfnamefont {F.}~\bibnamefont {Piazza}},\ and\ \bibinfo {author} {\bibfnamefont {T.}~\bibnamefont {Donner}},\ }\bibfield  {title} {\bibinfo {title} {Laser-painted cavity-mediated interactions in a quantum gas},\ }\href {https://doi.org/https://doi.org/10.1103/PRXQuantum.5.040332} {\bibfield  {journal} {\bibinfo  {journal} {PRX Quantum}\ }\textbf {\bibinfo {volume} {5}},\ \bibinfo {pages} {040332} (\bibinfo {year} {2024})}\BibitemShut {NoStop}%
\bibitem [{\citenamefont {{Baumgartner}}\ \emph {et~al.}(2024)\citenamefont {{Baumgartner}}, \citenamefont {{Pelliconi}}, \citenamefont {{Bandyopadhyay}}, \citenamefont {{Orsi}}, \citenamefont {{Sauerwein}}, \citenamefont {{Hauke}}, \citenamefont {{Brantut}},\ and\ \citenamefont {{Sonner}}}]{baumgartner:2024aa}%
  \BibitemOpen
  \bibfield  {author} {\bibinfo {author} {\bibfnamefont {R.}~\bibnamefont {{Baumgartner}}}, \bibinfo {author} {\bibfnamefont {P.}~\bibnamefont {{Pelliconi}}}, \bibinfo {author} {\bibfnamefont {S.}~\bibnamefont {{Bandyopadhyay}}}, \bibinfo {author} {\bibfnamefont {F.}~\bibnamefont {{Orsi}}}, \bibinfo {author} {\bibfnamefont {N.}~\bibnamefont {{Sauerwein}}}, \bibinfo {author} {\bibfnamefont {P.}~\bibnamefont {{Hauke}}}, \bibinfo {author} {\bibfnamefont {J.-P.}\ \bibnamefont {{Brantut}}},\ and\ \bibinfo {author} {\bibfnamefont {J.}~\bibnamefont {{Sonner}}},\ }\bibfield  {title} {\bibinfo {title} {{Quantum simulation of the Sachdev-Ye-Kitaev model using time-dependent disorder in optical cavities}},\ }\href {https://doi.org/https://doi.org/10.48550/arXiv.2411.17802} {\bibfield  {journal} {\bibinfo  {journal} {arXiv e-prints}\ ,\ \bibinfo {pages} {arXiv:2411.17802}} (\bibinfo {year} {2024})}\BibitemShut {NoStop}%
\bibitem [{dat(2026)}]{data}%
  \BibitemOpen
  \href {https://doi.org/10.5281/zenodo.18225513} {\bibinfo {title} {Data repository for this work}} (\bibinfo {year} {2026})\BibitemShut {NoStop}%
\bibitem [{\citenamefont {Roux}\ \emph {et~al.}(2021)\citenamefont {Roux}, \citenamefont {Helson}, \citenamefont {Konishi},\ and\ \citenamefont {Brantut}}]{roux_cavity-assisted_2021}%
  \BibitemOpen
  \bibfield  {author} {\bibinfo {author} {\bibfnamefont {K.}~\bibnamefont {Roux}}, \bibinfo {author} {\bibfnamefont {V.}~\bibnamefont {Helson}}, \bibinfo {author} {\bibfnamefont {H.}~\bibnamefont {Konishi}},\ and\ \bibinfo {author} {\bibfnamefont {J.-P.}\ \bibnamefont {Brantut}},\ }\bibfield  {title} {\bibinfo {title} {Cavity-assisted preparation and detection of a unitary {Fermi} gas},\ }\href {https://doi.org/10.1088/1367-2630/abeb91} {\bibfield  {journal} {\bibinfo  {journal} {New Journal of Physics}\ }\textbf {\bibinfo {volume} {23}},\ \bibinfo {pages} {043029} (\bibinfo {year} {2021})}\BibitemShut {NoStop}%
\bibitem [{\citenamefont {Birkl}\ \emph {et~al.}(1995)\citenamefont {Birkl}, \citenamefont {Gatzke}, \citenamefont {Deutsch}, \citenamefont {Rolston},\ and\ \citenamefont {Phillips}}]{birkl_bragg_1995}%
  \BibitemOpen
  \bibfield  {author} {\bibinfo {author} {\bibfnamefont {G.}~\bibnamefont {Birkl}}, \bibinfo {author} {\bibfnamefont {M.}~\bibnamefont {Gatzke}}, \bibinfo {author} {\bibfnamefont {I.~H.}\ \bibnamefont {Deutsch}}, \bibinfo {author} {\bibfnamefont {S.~L.}\ \bibnamefont {Rolston}},\ and\ \bibinfo {author} {\bibfnamefont {W.~D.}\ \bibnamefont {Phillips}},\ }\bibfield  {title} {\bibinfo {title} {Bragg {Scattering} from {Atoms} in {Optical} {Lattices}},\ }\href {https://doi.org/10.1103/PhysRevLett.75.2823} {\bibfield  {journal} {\bibinfo  {journal} {Physical Review Letters}\ }\textbf {\bibinfo {volume} {75}},\ \bibinfo {pages} {2823} (\bibinfo {year} {1995})}\BibitemShut {NoStop}%
\bibitem [{\citenamefont {Weidemüller}\ \emph {et~al.}(1995)\citenamefont {Weidemüller}, \citenamefont {Hemmerich}, \citenamefont {Görlitz}, \citenamefont {Esslinger},\ and\ \citenamefont {Hänsch}}]{weidemuller_bragg_1995}%
  \BibitemOpen
  \bibfield  {author} {\bibinfo {author} {\bibfnamefont {M.}~\bibnamefont {Weidemüller}}, \bibinfo {author} {\bibfnamefont {A.}~\bibnamefont {Hemmerich}}, \bibinfo {author} {\bibfnamefont {A.}~\bibnamefont {Görlitz}}, \bibinfo {author} {\bibfnamefont {T.}~\bibnamefont {Esslinger}},\ and\ \bibinfo {author} {\bibfnamefont {T.~W.}\ \bibnamefont {Hänsch}},\ }\bibfield  {title} {\bibinfo {title} {Bragg {Diffraction} in an {Atomic} {Lattice} {Bound} by {Light}},\ }\href {https://doi.org/10.1103/PhysRevLett.75.4583} {\bibfield  {journal} {\bibinfo  {journal} {Physical Review Letters}\ }\textbf {\bibinfo {volume} {75}},\ \bibinfo {pages} {4583} (\bibinfo {year} {1995})}\BibitemShut {NoStop}%
\bibitem [{\citenamefont {Miyake}\ \emph {et~al.}(2011)\citenamefont {Miyake}, \citenamefont {Siviloglou}, \citenamefont {Puentes}, \citenamefont {Pritchard}, \citenamefont {Ketterle},\ and\ \citenamefont {Weld}}]{miyake_bragg_2011-1}%
  \BibitemOpen
  \bibfield  {author} {\bibinfo {author} {\bibfnamefont {H.}~\bibnamefont {Miyake}}, \bibinfo {author} {\bibfnamefont {G.~A.}\ \bibnamefont {Siviloglou}}, \bibinfo {author} {\bibfnamefont {G.}~\bibnamefont {Puentes}}, \bibinfo {author} {\bibfnamefont {D.~E.}\ \bibnamefont {Pritchard}}, \bibinfo {author} {\bibfnamefont {W.}~\bibnamefont {Ketterle}},\ and\ \bibinfo {author} {\bibfnamefont {D.~M.}\ \bibnamefont {Weld}},\ }\bibfield  {title} {\bibinfo {title} {Bragg {Scattering} as a {Probe} of {Atomic} {Wavefunctions} and {Quantum} {Phase} {Transitions} in {Optical} {Lattices}},\ }\href {https://doi.org/10.1103/PhysRevLett.107.175302} {\bibfield  {journal} {\bibinfo  {journal} {Physical Review Letters}\ }\textbf {\bibinfo {volume} {107}},\ \bibinfo {pages} {175302} (\bibinfo {year} {2011})}\BibitemShut {NoStop}%
\bibitem [{\citenamefont {Hart}\ \emph {et~al.}(2015)\citenamefont {Hart}, \citenamefont {Duarte}, \citenamefont {Yang}, \citenamefont {Liu}, \citenamefont {Paiva}, \citenamefont {Khatami}, \citenamefont {Scalettar}, \citenamefont {Trivedi}, \citenamefont {Huse},\ and\ \citenamefont {Hulet}}]{hart_observation_2015}%
  \BibitemOpen
  \bibfield  {author} {\bibinfo {author} {\bibfnamefont {R.~A.}\ \bibnamefont {Hart}}, \bibinfo {author} {\bibfnamefont {P.~M.}\ \bibnamefont {Duarte}}, \bibinfo {author} {\bibfnamefont {T.-L.}\ \bibnamefont {Yang}}, \bibinfo {author} {\bibfnamefont {X.}~\bibnamefont {Liu}}, \bibinfo {author} {\bibfnamefont {T.}~\bibnamefont {Paiva}}, \bibinfo {author} {\bibfnamefont {E.}~\bibnamefont {Khatami}}, \bibinfo {author} {\bibfnamefont {R.~T.}\ \bibnamefont {Scalettar}}, \bibinfo {author} {\bibfnamefont {N.}~\bibnamefont {Trivedi}}, \bibinfo {author} {\bibfnamefont {D.~A.}\ \bibnamefont {Huse}},\ and\ \bibinfo {author} {\bibfnamefont {R.~G.}\ \bibnamefont {Hulet}},\ }\bibfield  {title} {\bibinfo {title} {Observation of antiferromagnetic correlations in the {Hubbard} model with ultracold atoms},\ }\href {https://doi.org/10.1038/nature14223} {\bibfield  {journal} {\bibinfo  {journal} {Nature}\ }\textbf {\bibinfo {volume} {519}},\ \bibinfo {pages} {211} (\bibinfo {year} {2015})}\BibitemShut {NoStop}%
\bibitem [{\citenamefont {Shao}\ \emph {et~al.}(2024)\citenamefont {Shao}, \citenamefont {Wang}, \citenamefont {Zhu}, \citenamefont {Zhu}, \citenamefont {Sun}, \citenamefont {Chen}, \citenamefont {Zhang}, \citenamefont {Fan}, \citenamefont {Deng}, \citenamefont {Yao}, \citenamefont {Chen},\ and\ \citenamefont {Pan}}]{shao_antiferromagnetic_2024}%
  \BibitemOpen
  \bibfield  {author} {\bibinfo {author} {\bibfnamefont {H.-J.}\ \bibnamefont {Shao}}, \bibinfo {author} {\bibfnamefont {Y.-X.}\ \bibnamefont {Wang}}, \bibinfo {author} {\bibfnamefont {D.-Z.}\ \bibnamefont {Zhu}}, \bibinfo {author} {\bibfnamefont {Y.-S.}\ \bibnamefont {Zhu}}, \bibinfo {author} {\bibfnamefont {H.-N.}\ \bibnamefont {Sun}}, \bibinfo {author} {\bibfnamefont {S.-Y.}\ \bibnamefont {Chen}}, \bibinfo {author} {\bibfnamefont {C.}~\bibnamefont {Zhang}}, \bibinfo {author} {\bibfnamefont {Z.-J.}\ \bibnamefont {Fan}}, \bibinfo {author} {\bibfnamefont {Y.}~\bibnamefont {Deng}}, \bibinfo {author} {\bibfnamefont {X.-C.}\ \bibnamefont {Yao}}, \bibinfo {author} {\bibfnamefont {Y.-A.}\ \bibnamefont {Chen}},\ and\ \bibinfo {author} {\bibfnamefont {J.-W.}\ \bibnamefont {Pan}},\ }\bibfield  {title} {\bibinfo {title} {Antiferromagnetic phase transition in a {3D} fermionic {Hubbard} model},\ }\href {https://doi.org/10.1038/s41586-024-07689-2} {\bibfield  {journal} {\bibinfo  {journal} {Nature}\ }\textbf {\bibinfo
  {volume} {632}},\ \bibinfo {pages} {267} (\bibinfo {year} {2024})}\BibitemShut {NoStop}%
\bibitem [{\citenamefont {Li}\ \emph {et~al.}(2017)\citenamefont {Li}, \citenamefont {Lee}, \citenamefont {Huang}, \citenamefont {Burchesky}, \citenamefont {Shteynas}, \citenamefont {Top}, \citenamefont {Jamison},\ and\ \citenamefont {Ketterle}}]{li_stripe_2017}%
  \BibitemOpen
  \bibfield  {author} {\bibinfo {author} {\bibfnamefont {J.-R.}\ \bibnamefont {Li}}, \bibinfo {author} {\bibfnamefont {J.}~\bibnamefont {Lee}}, \bibinfo {author} {\bibfnamefont {W.}~\bibnamefont {Huang}}, \bibinfo {author} {\bibfnamefont {S.}~\bibnamefont {Burchesky}}, \bibinfo {author} {\bibfnamefont {B.}~\bibnamefont {Shteynas}}, \bibinfo {author} {\bibfnamefont {F.~{\c C}.}\ \bibnamefont {Top}}, \bibinfo {author} {\bibfnamefont {A.~O.}\ \bibnamefont {Jamison}},\ and\ \bibinfo {author} {\bibfnamefont {W.}~\bibnamefont {Ketterle}},\ }\bibfield  {title} {\bibinfo {title} {A stripe phase with supersolid properties in spin–orbit-coupled {Bose}–{Einstein} condensates},\ }\href {https://doi.org/10.1038/nature21431} {\bibfield  {journal} {\bibinfo  {journal} {Nature}\ }\textbf {\bibinfo {volume} {543}},\ \bibinfo {pages} {91} (\bibinfo {year} {2017})}\BibitemShut {NoStop}%
\bibitem [{\citenamefont {Helson}\ \emph {et~al.}(2022)\citenamefont {Helson}, \citenamefont {Zwettler}, \citenamefont {Roux}, \citenamefont {Konishi}, \citenamefont {Uchino},\ and\ \citenamefont {Brantut}}]{HelsonOROASIFG2022}%
  \BibitemOpen
  \bibfield  {author} {\bibinfo {author} {\bibfnamefont {V.}~\bibnamefont {Helson}}, \bibinfo {author} {\bibfnamefont {T.}~\bibnamefont {Zwettler}}, \bibinfo {author} {\bibfnamefont {K.}~\bibnamefont {Roux}}, \bibinfo {author} {\bibfnamefont {H.}~\bibnamefont {Konishi}}, \bibinfo {author} {\bibfnamefont {S.}~\bibnamefont {Uchino}},\ and\ \bibinfo {author} {\bibfnamefont {J.-P.}\ \bibnamefont {Brantut}},\ }\bibfield  {title} {\bibinfo {title} {Optomechanical response of a strongly interacting fermi gas},\ }\href {https://doi.org/10.1103/PhysRevResearch.4.033199} {\bibfield  {journal} {\bibinfo  {journal} {Phys. Rev. Research}\ }\textbf {\bibinfo {volume} {4}},\ \bibinfo {pages} {033199} (\bibinfo {year} {2022})}\BibitemShut {NoStop}%
\end{thebibliography}%

\pagebreak
\clearpage
\renewcommand{\theequation}{S\arabic{equation}}
\renewcommand{\thefigure}{S\arabic{figure}}
\setcounter{figure}{0}

\section*{Supplemental Material}
\setcounter{page}{1}
\renewcommand{\thepage}{S\arabic{page}}
\subsection{Absorption imaging}

\subsubsection{Experimental setup and parameters}
We perform resonant, high-intensity absorption imaging using a \SI{1}{\micro\second} long pulse with an estimated intensity $I = 3.7\,I_{\text{sat}} =$ \SI{9.4}{\milli\watt\per\square\centi\meter}, with $I_{\text{sat}}$ the saturation intensity. 
 For the imaging pulse duration and intensity used, we expect a mean displacement of the atoms in the plane transverse to the imaging beam due to the random walk caused by the spontaneous emission events of \SI{0.35}{\micro\meter}.
 The objective used for imaging consists of a single aspheric lens (AFL50-60-U-B-285, asphericon) with a numerical aperture of $\text{NA} = 0.39$, which was tested to provide a measured resolution of $1.1 \upmu$m. We use an achromatic doublet lens to form a two-lens relay imaging setup and record images using a CMOS camera (ORCA-flash4.0), with a magnification of $12.0(3)$. The setup is aligned to the center of the atomic cloud coinciding with the center of the optical cavity.

\subsubsection{High intensity imaging calibration}

We use the modified Beer-Lambert law to account for the saturation of the imaging transition \cite{Reinaudi:2007aa, hueck_calibrating_2017,Horikoshi:2017ab}: 

\begin{equation}
    \frac{\text{d}I(x,y,z)}{\text{d}z} = -n(x,y,z)\sigma_{\text{eff}}\frac{1}{1+I(x,y,z)/I_{\text{eff}}}I(x,y,z)
    \label{Beer}
\end{equation}

where $I(x,y,z)$ denotes the total light intensity at position $(x,y,z)$ propagating along the $z$-direction, $n$ denotes atomic density, $\sigma_0 = 3\lambda_a^2/2\pi$ denotes the absorption cross-section, $I_{\text{eff}} = \alpha I_{\text{sat}}$ and $\sigma_{\text{eff}} = \sigma_0/\alpha$ denote the effective saturation intensity and effective cross-section. The parameter $\alpha$ captures experimental effects, as for example non-perfect polarization of the imaging beam, that reduce the cross-section compared to its theoretical expectation \cite{hueck_calibrating_2017}. In the experiment, the measured intensity is proportional to the count rate per pixel with position $(i,j)$ on the camera. Integrating equation (\ref{Beer}) along $z$ and expressing the intensity in terms of counts per pixel $C(i,j)$ yields \cite{hueck_calibrating_2017}:

\begin{equation}
     \frac{\sigma_0}{\alpha} n_{\text{2D}}(i,j) = -\rm{Ln}\left(\frac{C_{\text{out}}(i,j)}{C_{\text{in}}(i,j)} \right) + \frac{C_{\text{in}}(i,j)-C_{\text{out}}(i,j)}{C_{\text{eff}}}
    \label{OD}
\end{equation}

To extract the column density we calibrate the two parameters $C_{\text{eff}}$ and $\alpha$ independently. All the calibration procedures are performed in the absence of side pump or cavity field. For the parameter $C_{\text{eff}}$, following reference \cite{Horikoshi:2017ab}, we record optical densities using different imaging intensities and extract $C_{\text{eff}}$ from a linear fit to the logarithmic part of the right hand side of equation (\ref{OD}) as a function of its linear part, imposing that the OD remains constant for different imaging intensities. We calibrate the parameter $\alpha$ by comparing the atom number extracted from absorption imaging to the atom number extracted from cavity transmission spectroscopy, as described in \cite{roux_cavity-assisted_2021}.

\subsection{Suppression of coherent scattering in a thick grating}

\subsubsection{General formalism}

We now present a detailed mathematical treatment of imaging in our experiment to account for the finite thickness of the atomic cloud. We denote with $E(x,y,z)$ the field amplitude of the imaging beam, with wavevector along the $z$-direction $\mathbf{k} = k \, \mathbf{e}_z$, that propagates through an atomic cloud with density $n(x,y,z)$. We work in the framework of classical light scattering and model each atom as a scattering source that coherently emits from the imaging beam into spherical waves. The paraxial propagation of the electromagnetic field along the $z$-direction obeys:

\begin{align}
\label{PDE}
 \frac{\partial E_{k_x, k_y}(z)}{\partial z} = &\text{i}k_z(k_x,k_y)E_{k_x, k_y}(z) \\
 - &\int_{}^{}\int_{}^{} \frac{\sigma_0}{2(1+I(x,y,z)/I_{\text{sat}})} E(x,y,z) \notag \\
 & n(x,y,z)e^{-\I k_x x - \I k_y y}\D x\D y, \notag
\end{align}

where $k_z$ denotes the component of the scattered light along the $z$-direction $k_z = \sqrt{k^2-k_x^2-k_y^2}$ and $E_{k_x, k_y}(z)$ denotes the two dimensional Fourier transform of the electric field amplitude:

\[
E_{k_x, k_y}(z) = \int \int E(x,y,z) e^{-\I k_x x - \I k_y y} \D x \D y.
\]

For thin samples, we can neglect the difference between the $z$-component of the scattered wave vector $k_z$ and the total wave vector $k$ and approximate $k_z \approx k$. This is justified as long as we consider transverse angular frequencies $k_x,k_y$ such that the phase factor $\Delta \phi = (k-k_z)z_0$ is negligible, where $z_0$ denotes the characteristic thickness of the atomic cloud. This is the case if either the numerical aperture is small, or if the atomic cloud is thin along the $z$-direction. Then equation (\ref{PDE}) leads to the Beer-Lambert law (Eq. \ref{Beer}).

In our experiment however, the atomic cloud has a characteristic thickness $\sigma_z =$  $\SI{6}{\micro\meter}$, which yields the phase factor $\Delta \phi \approx$ $\SI{2.8}{\rad}$ and we can not justify the assumption $k_z \approx k$. 

Equation (\ref{PDE}) can nevertheless be solved in the regime of weak scattering, where we neglect the attenuation of the incoming beam through the cloud i.e. every atom is illuminated with the same amplitude $E_0$. The angular spectrum of the total scattered field is then:

\begin{equation}
\label{E_pert}
E_{k_x,k_y} = -\frac{\sigma_s E_0}{2}\int n_{k_x,k_y}(z') e^{\I (k-k_z)z'}\D z'
\end{equation}

where the integration runs over the whole atomic cloud. We further introduced the shortened the notation $\sigma_s := \frac{\sigma_0}{1+I/I_{\text{sat}}}$ and $n_{k_x,k_y}(z)$ denotes the two dimensional Fourier transform of the atomic density:
\[
n_{k_x, k_y}(z) = \int \int  n(x,y,z) e^{-\I k_x x - \I k_y y}\D x \D y.
\]

The electric field described by equation (\ref{E_pert}) then enters the imaging system. Within the approximations made, the measured Fourier transform $n_{\text{m}, k_x,k_y}$ of the atomic density recorded on the camera obeys:

\begin{equation}
\label{corr_pert}
n_{\text{m}, k_x,k_y} = \int n_{ k_x,k_y}(z') \cos((k-k_z)z')\D z'
\end{equation}

where $n_{\text{m}}(x,y) = \frac{1}{\sigma_s} \frac{I_0(x,y)-I(x,y)}{I_0(x,y)}$ and $I_0 \propto |E_0|^2$ is the incoming light intensity. Note that equation (\ref{corr_pert}) captures effects of atom-light scattering. Additional effects originating from the finite aperture of the imaging system can be taken into account via the modulation transfer function of the imaging system. The extracted density $n_{m}(x,y)$ as it is defined here coincides with the one extracted from the Beer-Lambert law in the two limit cases of either weak scattering or high intensity, such that $I \gg I_{\text{sat}}$. Dropping these assumptions complicates the analytical solution of equation (\ref{PDE}). However, as we will discuss in more detail at the end of this chapter, we expect that the basic principle of the thickness effect remains unchanged, especially in the regimes where the density modulations at high spatial frequencies are small.
\newline

\begin{figure}[ht!]
\includegraphics[width=0.48\textwidth]{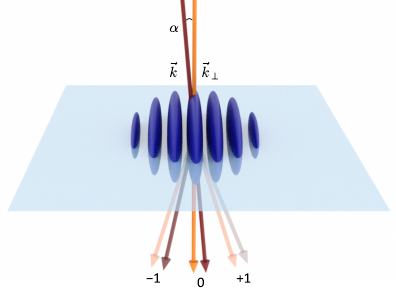}
\caption{Imaging of density-waves as scattering by a thick grating. We image the atomic cloud using an imaging beam propagating along $\mathbf{k}$. The plane in blue illustrates the plane defined by the wave vectors of the density modulation. The modulation along $\km$ leads to diffraction of the imaging light into the orders labeled with $0, \pm 1$. For an imaging beam orthogonal to the plane (orange), along $\mathbf{k}_\perp$, the diffraction efficiency is suppressed due to the thickness of the atomic cloud along the imaging beam direction. For an imaging beam incoming at a finite angle $\alpha$ with respect to  $\mathbf{k}_\perp$ (brown) the diffraction into one of the first orders is favored which leads to an overall increase of the scattering efficiency. Drawn is the experimental condition with $\alpha =  5.2(9)$°.
\label{fig5}}
\end{figure}

\subsubsection{Coherent suppression factor}

To progress analytically, we model the atomic density with a Gaussian envelope and a density modulation at a single frequency $\km$:
\begin{multline}
n(x,y,z) = e^{-\frac{x^2}{2 \sigma_x^2}-\frac{y^2}{2 \sigma_y^2} - \frac{z^2}{2 \sigma_z^2}}\times \\ \left( n_0 + n_1 \cos(k_{-} (x \sin(\theta) + y \cos\theta) \cos\alpha + k_{-}z\sin\alpha))\right)
\end{multline}
where $n_0$ is the peak density, $n_1$ is the modulation amplitude and $\theta$ parametrizes the direction of $\km$ in the $x,y$ plane. Here $\alpha$ represents the angle of the imaging beam with respect to the $x,y$ (pump-cavity) plane, deviating from being orthogonal to $\km$ (see Fig. \ref{fig5}). We assume the plane of image formation, i.e. the front facet of the objective lens and the camera, to be orthogonal to the imaging beam direction. According to equation (\ref{corr_pert}), the Fourier amplitude of the density image recorded on the camera is then:

\begin{align}
n_{\text{m}, k_{-,x} k_{-,y}} &= \frac{1}{2}(n_{k_{-,x} k_{-,y} (k-k_z)} + n_{k_{-,x} k_{-,y} -(k-k_z)}) \\
&\approx \frac{1}{2} \bigl(e^{-\frac{\sigma_z^2}{2}(k - k_{z} + k_{-,z})^2} + e^{-\frac{\sigma_z^2}{2}(k - k_{z} - k_{-,z})^2} \bigr) n_{k_{-,x}k_{-,y}, k_{-,z}},
\end{align}

where we introduced $k_{-,z} = k_{-}\sin(\alpha)$ and 
\[
n_{k_{-,x}k_{-,y}, k_{-,z}} = \int \int \int n(x,y,z) e^{-\I k_{-,x} x - \I k_{-,y} y - \I k_{-,z} z} \D x \D y \D z
\]
is the Fourier transform of the atomic density evaluated at $\km$. The prefactor in front of the real Fourier component represents the effect of coherent suppression of diffraction due to the finite cloud thickness. The factor is unity for $\sigma_z\to0$, and has a local maximum when $\alpha$ verifies the Bragg condition. Note that in the experiment, we observe two sharp peaks at $\km$ which fulfill $\delta k_{-}/k_{-} < 0.1$, $\delta k_{-}$ denoting the full width at half maximum of the observed peak, which in turn translates to a difference in the phase factor $\Delta \phi$ of less than $10\%$. This justifies the assumption of a constant suppression factor in the vicinity of $\km$ and the interpretation of $\hatnmpk^2$ in terms of density correlations. 

Our experiment is conceptually similar to other scattering experiments, where the structure factor is extracted through coherent Bragg scattering \cite{birkl_bragg_1995, weidemuller_bragg_1995, miyake_bragg_2011-1, hart_observation_2015, shao_antiferromagnetic_2024, li_stripe_2017}, with the difference that we recombine the Bragg scattered light with the zeroth order, enabling the extraction of phase information, equivalent to a holographic read out.

In our experiment we estimate $\alpha \approx \SI{5.2(9)}{\degree}$, leading to a total suppression factor of $\approx 0.24(8)$. The $+1$ order is entirely suppressed while the $-1$ order is reduced by about a factor of two. We verified the strong asymmetry by directly imaging the back Fourier plane of the objective onto the camera, where only one order of diffraction was visible.

The experimental parameters have been optimized for signal to noise ratio in absorption imaging, and the weak scattering assumption leading to Eq. \ref{corr_pert} are not fulfilled. A situation closer to the experimental condition consists in considering a weak modulation amplitude $n_1$, leading to a weak diffracted field instead of a weak total absorption. No analytic solution can be obtained in this case, but the suppression factor keeps the same shape up to an overall correction originating from the changes of the saturation parameter along $z$, slightly reducing the suppression factor. 

\subsection{Theory and Hamiltonian with phase correlations}

In this section we review the mean-field description of the density-wave ordering transition (see \cite{helson:2023aa} for example) keeping  track explicitly of the different phases of both the cavity field and of the atomic density waves. This allows us to identify the different contributions to the photonic signal, and elucidate the role of the spontaneous symmetry breaking. 

A retro-reflected laser beam pumps the atoms from the side at angle $\theta$ from the cavity axis. The pump and its retroreflection have parallel polarization along $\bm{\epsilon}_{z}$, the axis of the magnetic bias field which is also orthogonal to the (pump-cavity) plane. Its frequency is set in the vicinity of a longitudinal mode of the cavity with annihilation operator $\hat{a}$. The cavity mode is either TEM$_{00}$ or TEM$_{01}$, with a characteristic spatial dependence of the mode field $\uc(\R)$. We index the two running waves of the side pump by the sign of their wavevector $\pm \kp$ and denote with $\up(\R)$ the spatial dependence of the pump beam.

The total field $\bm{\Phi}_\text{tot}$ reads:
\begin{equation}
  \hat{\Phi}_\text{tot}(\R) = g \uc(\R) \cos(\kc \R)  \bm{\epsilon}_{z} \hat{a} + \sum_\pm \frac{\Omega_\pm}{2} \up(\R) \E^{\pm \I \kp r}  \E^{\I \varphi_\pm} \bm{\epsilon}_{z}, 
\end{equation}

where $g$ and $\Omega$ denote the maximum single atom-photon coupling strength and the maximum Rabi frequency of the pump field respectively. In the following, we assume that the pump and its retroreflection have equal Rabi frequencies $\Omega = \Omega_+ = \Omega_-$ (assumption of perfect retroreflection). For simplicity, we consider the TEM$_{00}$ mode, such that the spatial dependence of the mode field is a Gaussian, and we approximate it by $\uc(\R) \approx 1$ since the size of the cloud is small compared to the cavity mode waist and similarly we assume $\up(\R) \approx 1$.  Without loss of generality, we have chosen here a longitudinal mode with an even number of nodes, corresponding to the cosine function in the equation, where $\R=0$ corresponds to the cavity center.

The Hamiltonian describing the cavity field and the light-matter interaction then writes:
\begin{equation}
\begin{aligned}
  \hat{H} & = -{\Delta}_c \hat{a}^\dagger \hat{a} + \frac{1}{\Delta_a} \int \D \R \hat{n}(\R) \hat{\Phi}_\text{tot}^\dagger \hat{\Phi}_\text{tot},
\end{aligned} 
\end{equation}
within the rotating wave approximation at the drive frequency and in the dispersive regime, after elimination of the atomic excited state at detuning $\Delta_a$ from the side pump. Here we denote with $\hat{n}(\R)$ the total density operator. 

We get the three terms of the Hamiltonian:
\begin{equation}
\begin{aligned}
  \hat{H}_{\text{ph}} & = -\tilde{\Delta}_c \hat{a}^\dagger \hat{a} \\
  \hat{H}_{\text{p}} & = \frac{\Omega^2}{\Delta_a} \int \D \R \hat{n}(\R) \cos^2(\kp \R + \frac{\varphi_+ - \varphi_-}{2})  \\
  \hat{H}_{\text{lm}} & = \eta_0 \left(\hat{a}^\dagger \E^{\I \frac{\varphi_+ + \varphi_-}{2}} + \hat{a} \E^{-\I \frac{\varphi_+ + \varphi_-}{2}} \right) \hat{\Theta},
\end{aligned} 
\label{three_parts_hamiltonian}
\end{equation}

where $\eta_0 = \Omega g/\Delta_a$, $\tilde{\Delta}_{\text{c}} = \Delta_c - g^2/\Delta_a \int \D \R \hat{n}(\R) \cos^2(\kc \R)$ is the detuning of the pump to the dispersively-shifted cavity resonance and $\hat{\Theta}$ is the atomic order parameter:

\begin{equation}
\begin{aligned}
  \hat{\Theta} & = \frac{1}{2} \int \D \R \hat{n}(\R) \left(\cos( \kplus \R - \frac{\varphi_- - \varphi_+}{2}) \right. \\
  & \quad\quad\quad\quad\quad\quad \left.  + \cos(\km \R - \frac{\varphi_- - \varphi_+}{2}) \right) \\
  & = \frac{1}{4} \left(\hat{n}_{\kplus} \E^{-\I\frac{\varphi_- - \varphi_+}{2}} + \hat{n}_{\kplus}^\dagger \E^{\I\frac{\varphi_- - \varphi_+}{2}} \right. \\
  & \quad\quad \left. + \hat{n}_{\km} \E^{-\I\frac{\varphi_- - \varphi_+}{2}} + \hat{n}_{\km}^\dagger \E^{\I\frac{\varphi_- - \varphi_+}{2}}\right) \\
\end{aligned} 
\label{def_order_param}
\end{equation}

using $\hat{n}_{\mathbf{q}} = \int \D \R \hat{n}(\R) \E^{\I \mathbf{q} \R}$. The atomic order parameter is real-valued. It is non-zero in the case of a modulation of the density with wavevector either $\km$ or $\kplus$ with a phase shifted by $\frac{\varphi_- - \varphi_+}{2}$ from the origin $\R=0$. 
Note that the Hamiltonian is invariant under the transformation $\left(\hat{a}, \hat{\Theta}\right) \rightarrow \left(\hat{a} \E^{\I \pi}, -\hat{\Theta}\right)$.

In the regime of large detuning $\Delta_c \gg \kappa$, we neglect the dissipation due to the cavity decay at a rate fixed by the cavity linewidth $\kappa$. From the Heisenberg equation of motion for the operator $\hat{a}$ together with the assumption that the cavity field adiabatically follows the atoms, its steady state value can be expressed in terms of the atomic order parameter: 

\begin{equation}
\label{linear-eq}
\begin{aligned} 
  \hat{a} \approx \E^{\I \frac{\varphi_+ + \varphi_-}{2}} \frac{\eta_0}{\tilde{\Delta}_c} \hat{\Theta}. \\
\end{aligned} 
\end{equation}

This corresponds to the known operator identity relating the cavity field operator and $\hat{\Theta}$, where the role of the pump phases is explicitly visible. \newline
 
Since $\hat{\Theta}$ is real-valued, its sign corresponds to the $\mathbb{Z}_2$ symmetry breaking, which directly translates into a phase shift of $\phi_\text{symm}=0$ or $\pi$ for the phase of the density-wave $\varphi_\text{DW}$ (or equivalently a shift of half the period of the density-wave) and simultaneously for the phase  of the field. From the equations (\ref{three_parts_hamiltonian}) and (\ref{def_order_param}) we get the following relations for the different phases we expect to hold for the measurement outcomes:

\begin{equation}
\begin{aligned} 
  \varphi_a & = \frac{\varphi_+ + \varphi_-}{2} + \phi_\text{symm} \\
  \varphi_\text{DW} & = \frac{\varphi_- - \varphi_+}{2} + \phi_\text{symm} \\
\end{aligned} 
\end{equation}

If we consider the difference between these two equations, we suppress the dependence on the phase of the symmetry breaking and only retain the dependence with a single phase:

\begin{equation}
\begin{aligned} 
  \varphi_a - \varphi_\text{DW} & = \varphi_+,\\
\end{aligned} 
\end{equation}

which supports our measurements presented in Fig.~\ref{fig3}(f). The same reasoning applies to the TEM$_{01}$ mode and taking into account the spatial density of the atomic gas, as their length scales are much larger than the light wavelength. The atomic order parameter acquires a spatial structure from the mode field $\uc(\R)$ and the spatial density, and the steady-state equation (\ref{linear-eq}) between the cavity field and the order parameter holds.

\subsection{Heterodyne detection}

\begin{figure}[t]
\includegraphics[width=0.48\textwidth] {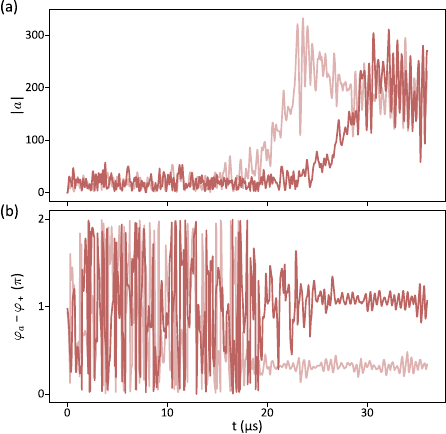}
\caption{
Examples of cavity fields measured by heterodyne detection. (a),(b) Two examples of demodulated beatnotes, in light red and dark red (amplitude and phase in (a) and (b) respectively), for the same experimental protocol of abruptly turning on the pump potential to $V_0 = 18.0(4)V_\text{0c}$ ($\tilde{\Delta}_c/2\pi = \SI{-2.7(1)}{\MHz}$,  $V_0 = 2.02(3) \Erk$, $N = 5.2(3) \times 10^5$). The phase is referenced to the phase of the pump beam, obtained from an additional heterodyne detection setup with common local oscillator. 
\label{fig:heterodyne}}
\end{figure}

We measure the phase of the cavity field using heterodyne measurements. We combine the field emitted by the cavity with a local oscillator with a total power of $\SI{2}{\mW}$ detuned by $\SI{60}{\MHz}$ on a polarising beam splitter. The detuning is chosen to be much larger than the cavity linewidth $\kappa/2\pi = \SI{77}{\kHz}$ to ensure that no back-scattered light from the local oscillator couples into the cavity. The signal beam and the local oscillator are then coupled to a fiber with orthogonal polarizations. Using a waveplate and a polarising beam splitter, we beat the two beams and detect them on the two output ports of the splitter using a balanced photodetector (Femto HBPR-100M-60K-SI-FST). The difference signal is then amplified, down-converted to $\SI{5}{\MHz}$ and digitally sampled at a rate of $\SI{32}{\MHz}$.
The beating frequency of $\SI{5}{\MHz}$ is chosen as a compromise between technical limits in hard disk space and the resolution of the fastest timescale in ordering dynamics of the system at the $\si{\micro\second}$ level.

We compute the power spectral density of the beating signal and observe non-zero signal amplitude at the expected beating frequency of $\SI{5}{\MHz}$ when the power of the pump exceeds the critical power of the phase transition. We use IQ demodulation, with a low-pass filter of $\SI{1.25}{\MHz}$ for most of the data presented in this work, to extract the amplitude and phase of the leakage field at the beating frequency. The conversion from a voltage to a number of intracavity photons relies on a calibration using a single photon counting module and a previous calibration of its detection efficiency in our earlier work \cite{HelsonOROASIFG2022}.

As described in the previous paragraph, the phase of the cavity field is locked to the phases of the pump beam and its retro reflection. For this reason we record the phase of the pump $\varphi_{+}$ with an additional heterodyne detection setup based on a fast photodiode. Both measurements, of pump and cavity fields, are carried out outside of the vacuum chamber. The fields pass similar optics on the same breadboard, before they get guided to their respective heterodyne setups, which suppresses common mode phase noise due to the mechanical motion of the optical breadboard. By using the same local oscillator for both heterodyne detections, we further suppress the drift of the phase of the local oscillator itself. We therefore faithfully measure the difference $\varphi_a - \varphi_+$.  

Examples of the time evolution of the recorded amplitude and phase of the cavity field for experimental protocols where pump potential is abruptly turned on above the critical point are shown in Fig.~\ref{fig:heterodyne}.

\subsection{Threshold extraction}
We numerically extract the critical pump lattice depth $V_{0\text{c}}$ based on the cavity field measurement for both experimental protocols, quenching and linearly ramping up $V_{0}$. In the case of a linear ramp, we extract $V_{0\mathrm{c}}$ over a single experimental realization using the fit function 
\begin{equation}
    |a|^2(V_0) = \theta(V_0 - V_{0\mathrm{c}})\times m\; \times (V_0 - V_{0\mathrm{c}}),
\end{equation}
with $\theta(V_0 - V_{0\mathrm{c}})$ the Heaviside function and the fit parameters $V_{0\mathrm{c}}$ as well as the slope of the initial onset of the detected cavity field intensity $m$, similar as in our previous work \cite{zwettler:2025ac}. For the quench protocol we use the same fitting function, fitting to the total detected photon number over several experimental repetitions with varying $V_0$. For a more detailed description, please refer to our previous work \cite{zwettler:2025ab}.
In order to compare the threshold extracted via the photon measurement to the in-situ density wave measurement, as it is shown in Fig. \ref{fig2} (e), we extract the cavity field endpoints immediately before we take the absorption image. We then extract $V_{0\mathrm{c}}$ as the crossing point of the cavity field endpoints with a threshold set above the noise floor, verifying that the exact threshold value does not influence the conclusions drawn. This method is used as an alternative to the fit with a linear function as it has been proven to be robust in cases with a limited sampling of the signal onset.

\subsection{Atom-photon correlations for varying parameters}

Our analysis of the atom-photon correlations is directly related to the Heisenberg equation for the field, which connects the atomic order parameter to the cavity field in the steady-state (\ref{linear-eq}). The detuning of the pump beam relative to the dispersively shifted cavity $\tilde{\Delta}_c = \Delta_c - \delta$ is calculated by measuring the dispersive shift of the cavity just before quenching the long-range interactions. The parameter $\eta_0 = \Omega g/\Delta_a$ is deduced from a calibration of the potential $V_0 = \Omega^2/\Delta_a$ created by the pump using Kapitza-Dirac diffraction of the atoms in the pump lattice and by measuring $\delta$. We do not take into account possible losses due to the pump beams that would decrease the dispersive shift (increase the detuning) nor the organization of the atoms along the cavity direction in the density-wave ordered phase that would increase the dispersive shift (decrease the detuning) compared to the unmodulated phase with an overlap of $0.5$ with the cavity lattice. Equation (\ref{linear-eq}) implies that the atom-photon correlations observed in the quench protocol hold for all systems and for all circumstances. We further test this hypothesis by repeating our measurements varying different experimental parameters. 

\subsubsection{BEC-BCS crossover}

We vary the bias magnetic field in the vicinity of the broad Feshbach resonance for the two lowest hyperfine states of Lithium atoms at $\SI{832}{\Gauss}$ in order to tune their scattering properties. It modifies the short-range properties of the atomic gas, including its compressibility, which dictates the value of the critical pump potential $V_{0\text{c}}$ to reach density-wave ordering \cite{helson:2023aa}. We study the correlations between the atomic order parameter and the amplitude of the cavity field by quenching the system above threshold, with the same normalized pump potential $V_0 / V_{0\text{c}}$ and at fixed $\tilde{\Delta}_c$, for three choices of the short-range interaction parameter.

\begin{figure}[ht!]
\includegraphics[width=0.48\textwidth]{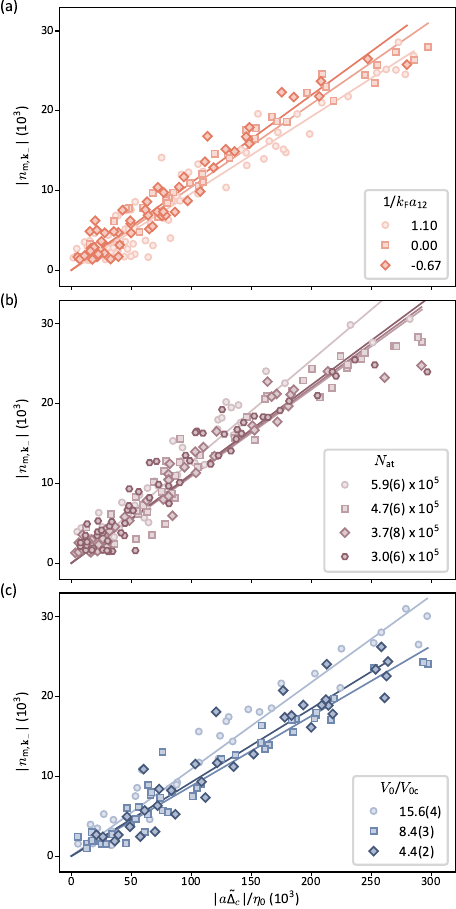} 
\caption{Atom-photon correlations for varying parameters. (a) Varying interaction parameters at fixed atom number $N = 5.0(6) \times 10^5$ and pump potential $V_0 = 13.8(3)V_{0\text{c}}$. (b) Varying the number of atoms at fixed effective pump-cavity detuning $\tilde{\Delta}_c/2\pi = \SI{-2.6(2)}{\MHz}$ and pump lattice depth $V_0 = 2.11(3)\Erk$
, for a unitary Fermi gas. (c) Varying the detuning and pump potential at fixed atom number $N = 5.2(5) \times 10^5$. $V_0 = 15.6(4)V_{0\text{c}}$ ($\tilde{\Delta}_c/2\pi = \SI{-3.2(1)}{\MHz}$,  $V_0 = 2.16(3) \Erk$), $V_0 = 8.4(3)V_{0\text{c}}$ ($\tilde{\Delta}_c/2\pi = \SI{-2.6(1)}{\MHz}$,  $V_0 = 0.97(1) \Erk$), and $V_0 = 4.4(2)V_{0\text{c}}$ ($\tilde{\Delta}_c/2\pi = \SI{-2.6(1)}{\MHz}$,  $V_0 = 0.51(1) \Erk$). The solid lines represent linear fits to the data.
\label{fig:supp_UFG}}
\end{figure}

\subsubsection{Varying long-range interaction strength and atom number}

In the unitary limit we change the number of atoms, while keeping $V_0$ and $\tilde{\Delta}_c$ constant. Further, we vary $V_0/V_{0\text{c}}$, keeping atom number constant. The linear relation, once the pump and atomic parameters $\eta_0$ and $\tilde{\Delta}_c$ are taken into account, is observed with a constant slope within the detection noise.

\subsection{Atom-photon correlations at long times}

\begin{figure}[t]
\includegraphics[width=0.48\textwidth]{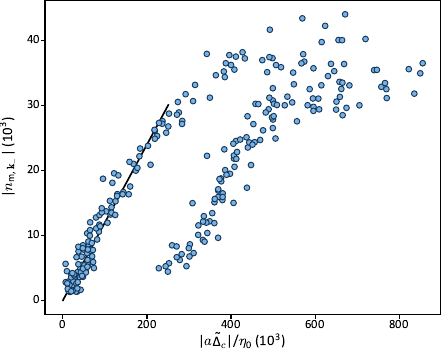} 
\caption{Atom-photon correlations up to longer times following a quench, compared to Fig.~\ref{fig3}. The black line is a linear fit to the data in the range shown in Fig.~\ref{fig3}.
\label{fig:supp_lin_rel}}
\end{figure}

At long times (time scales longer than the rise time of the cavity field), we observe deviations from the linear correlation between the measured atomic order parameter and the cavity field, as it is shown in Fig. \ref{fig:supp_lin_rel}. 

The linear coefficient changes by more than a factor $2$. We attribute this variation to three main mechanisms: (i) in the presence of a density wave, the cloud size changes along the axis of the microscope, inducing finite thickness-related contrast variations, (ii) the atoms move during the $\SI{1}{\micro\second}$-long imaging pulse due to the release of potential energy, which reduces the contrast of the density modulation at large cavity fields and (iii) $\Theta$ comprises both the $\mathbf{k}_{-}$ and $\mathbf{k}_{+}$ contributions. Close to the transition, ordering is dominated by the lower energy mode at $\mathbf{k}_{-}$ and the microscope images faithfully reflect the density waves, but deeper in the ordered phase, previous works show that the $\mathbf{k}_{+}$ component contributes significantly to the photon field dynamics in the cavity \cite{zwettler:2025ab,marijanovic:2024aa}, but it is not detected in the images. An additional contribution, however small, comes from the change of $\tilde{\Delta}_c$ as the system orders, due to the change in the refractive index of the atomic cloud in the cavity.

\subsection{Local analysis of amplitude and phase in the TEM$_{00}$ mode}

\begin{figure}[t]
\includegraphics[width=0.48\textwidth]{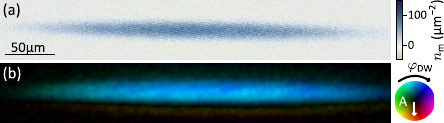} 
\caption{Local analysis of absorption images of density-waves in the TEM$_{00}$ mode of the optical cavity. (a) Example of a single shot absorption image. The image is taken following an instantaneous quench of $V_0$ up to $V_0 = 24 V_{0 \text{c}}$ and an evolution time of $\SI{35}{\micro\s}$. (b) Local amplitude and phase of the density wave at $\km$. The amplitude is encoded in the brightness and the phase is represented by the color scale. Shown is the average over $100$ experimental repetitions.
\label{fig:TEM00_local}}
\end{figure}

We analyze locally the recorded absorption images for observed density-wave ordering of the atomic cloud in the TEM$_{00}$ mode. For this we extract the Fourier component at $\km$ with a spatial resolution of $\SI{3.4}{\micro\meter}$. The results of the extracted amplitude and phase, averaged over $100$ experimental repetitions, are shown in Fig.~\ref{fig:TEM00_local}(b), with one example of a corresponding, single-shot image shown in Fig.~\ref{fig:TEM00_local}(a). We observe an amplitude envelope of the density modulation that is well-described by a Gaussian profile, closely matching the shape observed in the atomic density. Furthermore, we observe a constant phase over the extent of the atomic cloud, as it is visible encoded in the color scale in Fig.~\ref{fig:TEM00_local}(b). This supports the interpretation of the recorded images as a single-mode density-wave extending over the whole atomic cloud. 

\subsection{Coupling to the TEM$_{01}$ mode}

\begin{figure}[h]
\includegraphics[width=0.48\textwidth]{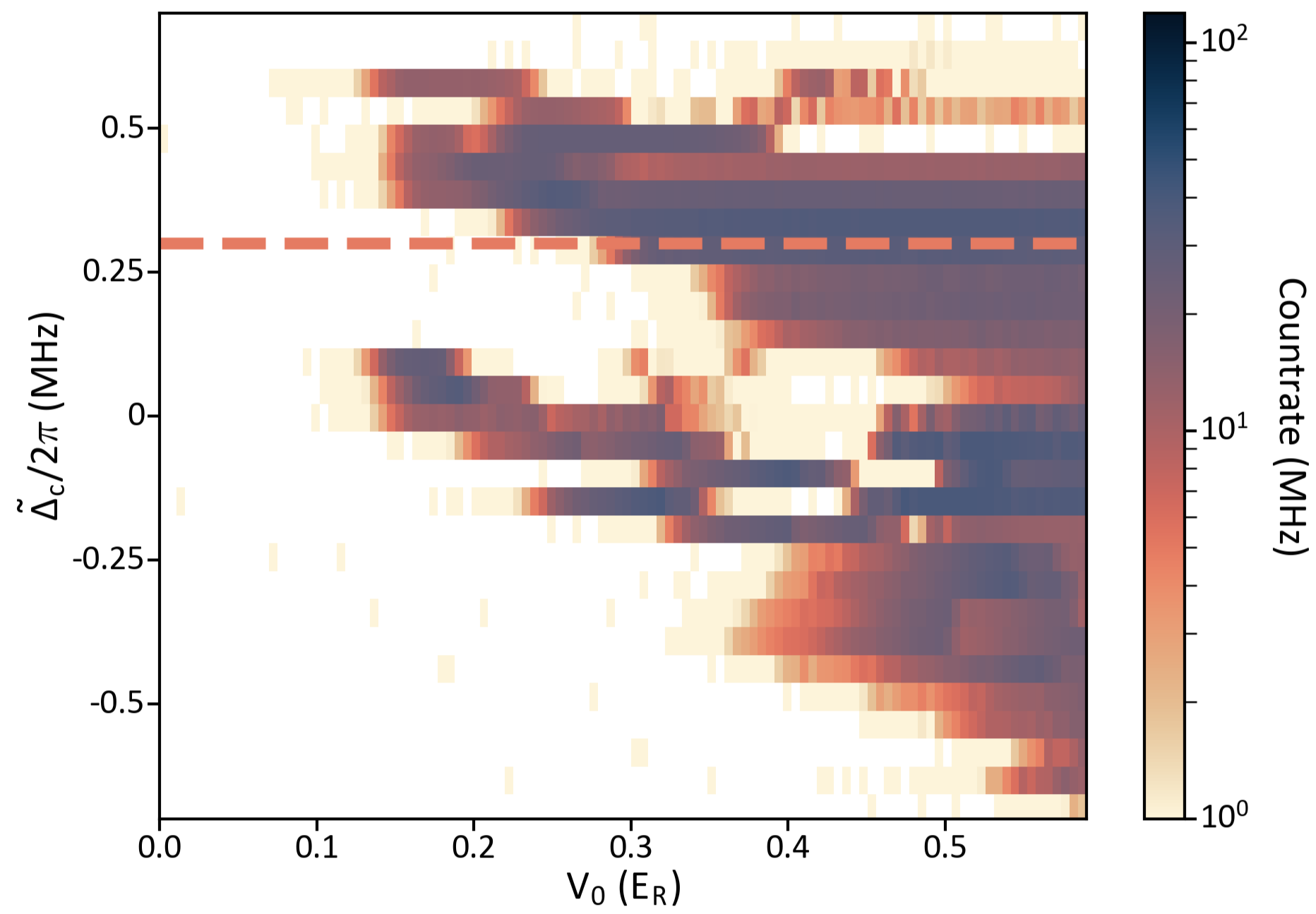} 
\caption{Photon rate emitted by the cavity as a function of the pump power $V_0$ and pump-cavity detuning for a pump laser frequency tuned close to the TEM$_{01}$ and TEM$_{10}$ modes. Owing to the cavity’s intrinsic astigmatism, two distinct phase boundaries appear, each corresponding to resonance with one of the modes. The dashed line marks the conditions under which the data in Fig.~\ref{fig4} were taken. 
\label{fig:TEM01}}
\end{figure}

When operating with the pump laser frequency close to the TEM$_{01}$ and TEM$_{10}$ modes we observe self-ordering in the form of superradiant light emission in the cavity for an applied pump potential above a critical value. Fig. \ref{fig:TEM01} shows the phase diagram in the detuning-pump potential plane. We observe two distinct corner-shaped phase boundaries corresponding to the two first-order modes, separated in frequency by \SI{0.5}{\MHz} 
due to astigmatism in the cavity. The dashed horizontal line indicates the detuning at which the data presented in Fig.~\ref{fig4} were acquired.

\end{document}